\documentclass[a4paper]{tufte-handout}

\usepackage{amsmath}
\usepackage{greek}
\usepackage{mythue}

\title{Thue's 1914 paper: a translation}

\author{James F. Power\marginnote{Department of Computer Science,
National University of Ireland, Maynooth, 
Co. Kildare, Ireland.}}

\begin{document}

\maketitle

\section{Introduction}

Axel Thue published four papers directly relating to the theory of words
and languages:
\begin{itemize}
\item two on patterns in infinite strings in 1906 and 1912
\cite{thue06}\cite{thue12}
\item and two on the more general problem of
transformations in 1910 and 1914 \cite{thue10}\cite{thue14} 
\end{itemize}
Both the 1906 and 1912 papers have been translated and
discussed extensively by Jean Berstel \citep{berstel95}, and are known,
among other contributions, for their presentation of the Thue-Morse sequence.
Thue's 1910 paper deals with transformations between trees, and is thus
a more direct predecessor of his 1914 paper.
It been discussed by Steinby and Thomas \citep{steinby00}.

\bigskip

These notes are intended to accompany a reading of Thue's 1914 paper,
which has not hitherto been discussed in detail.  Thue's paper is mainly
famous for proving an early example of an undecidable problem, cited
prominently by Post \citep{post47}.  However, Post's paper principally
makes use of the definition of \emph{Thue systems}, described on
the first two pages of Thue's paper, and does not depend on the
more specific results in the remainder of Thue's paper.

\bigskip

Thus, Thue's paper has been ``passed by reference'' into the history of
computing, based mainly on a small section of that work.  A closer
study of the remaining parts of that paper highlight a number of
important themes in the history of computing: the transition from
algebra to formal language theory, the analysis of the
"computational power" (in a pre-1936 sense) of rules, and the
development of algorithms to generate rule-sets.

\bigskip

\paragraph{Structure:}
This document is in three sections
\begin{itemize}
\item We present a brief overview of Thue's paper and a motivation for
  studying it (pp. 2-3).  This is an extended abstract of a talk to
  be presented at the \emph{International Conference on the History and
  Philosophy of Computing} (HaPoC 2013), 28-31 October, 2013, Ecole
  Normale Superieure, Paris.

\item We provide some notes on the contents of the paper (pp. 4-10),
  which are intended to be read in conjunction with the paper (or its
  translation).

\item The last section is a translation of Thue's paper, numbered as in
  Thue's \emph{Selected Papers} \citep{nagell77}, pages 493-524.

\end{itemize}

\newpage
\section{An overview of Thue's paper}

\medskip

Rarely has any paper in the history of computing been given
such a prestigious introduction as that given to Axel Thue's paper by Emil
Post in 1947 \citep{post47}:

\begin{quotation}
``Alonzo Church suggested to the writer that a certain problem of Thue
\citep{thue14} might be proved unsolvable ...''
\end{quotation}

However, only the first two pages of Thue's paper are directly relevant to
Post's proof, and, in this abstract, I hope to shed some light on the
remaining part, and to advocate its relevance for the history of
computing.

\paragraph{Thue Systems}
Thue's 1914 paper is the last of four he published that directly
relate to the theory of words and languages
\citep{berstel95,steinby00}.  In this 1914 paper, 
Thue introduces a system consisting of pairs of corresponding strings
over a fixed alphabet:  
$$\begin{array}{ccccc}
A_1,& A_2,& A_3,& \ldots,& A_n\\
B_1,& B_2,& B_3,& \ldots,& B_n,
\end{array}$$
and poses the problem: given two arbitrary strings $P$
and $Q$, can we get one from the other by replacing some substring
$A_i$ or $B_i$ by its corresponding string?  Post called these systems
of ``Thue type'' and proved this problem to be recursively unsolvable.

\paragraph{Reception of Thue's Work}

Thue's earlier work was not widely cited but often rediscovered
independently \citep{hedlund67}, and something similar seems to have
happened with the 1914 paper.

For example, Thue is not among the 547 authors in Church's 1936
\emph{Bibliography of Symbolic Logic} \citep{church36}, nor is Thue cited in Post's
major work on tag systems, correspondence systems, or normal systems
before 1947 \citep{post43,post46}.  His work appears to have had no
direct influence on the development of formal grammars by Chomsky in
the 1950s \citep{chomsky59,scholz07}.  Most subsequent references to
Thue's paper (where they exist) note it only for providing a
definition of Thue systems.

\paragraph{Thue's awareness}

Thue explicitly understood the general meta-mathematical context (that
we now associate with \emph{Hilbert's programme}), describing the
problem as being of relevance to one of the ``most fundamental
problems that can be posed''.  

Further, he phrases the problem in terms that have become quite
familiar in the post-1936 world: 
\begin{quotation}
``... to find a method, where
one can always calculate in a predictable number of operations, ...''
\end{quotation}
This language parallels that used in Hilbert's 10th problem in
1900 \citep{hilbert02}, and places Thue's work firmly in what we would
now regard as computing, rather than pure algebra.

\paragraph{Foundations of Language Theory}

Having posed the general problem in \S II of his paper, Thue then
presents an early example of a proof of (what we would now call)
\emph{termination} and \emph{local confluence} for a system where the
rules are non-overlapping and non-increasing in size.

When reducing some string $P$, we must
find some occurrence of $A_i$ and replace it with $B_i$.  A difficulty
arises if there is an overlap: some substring $CUD$ in $P$, such that
$A_i$ matches both $CU$ and $UD$, and thus choosing one option will
eliminate our ability to later choose the other.

In \S IV, Thue presents the string $U$ as a \emph{common divisor} of
$CU$ and $UD$ and then shows how we can apply \textbf{Euclid's
  algorithm} to derive a Thue system from this.  Euclid's algorithm
had been considerably generalised throughout the 19th century, but
here the string $U$ ``measures'' the strings $CU$ and $UD$ just as
Euclid's lines measure each other (\emph{Elements}, Book 10,
proposition 3).

Thue derives another \emph{algorithm} in \S V which, given two strings
$P$ and $Q$ will derive those strings equivalent to them, and
gradually reduce them to a core set of irreducible strings, providing
a solution to the word problem in a restricted case.  He investigates
variants of these presentations \emph{based on their syntactic
  properties} in \S VI and gives some examples in \S VII.

We remark that from the identity $CU \equiv UD$ we can derive rules
of the form $CU \rightarrow UD$, and that this template is precisely what Post
termed \emph{normal form} for his rewriting systems.

\paragraph{Thue's ``completion'' algorithm}

In \S VIII of his paper Thue develops an algorithm to derive
a system of equations from any given sequence $R$.  This is
interesting not just for its structure (the algorithm iterates until
it reaches a fixed point) but also for its use of overlapping
sequences as a generation mechanism.  

Starting from some given identity sequence $R$ we can identify all
pairs where $R \equiv CU \equiv UD$, and then add the rules $C \leftrightarrow D$
to the Thue system.  We can then apply these rules using $R$ as a
starting symbol to derive a further set of identity sequences $R_1,
R_2, \ldots$.  These, in turn, can be factored based on overlaps to
provide a further set of rules $C_i \leftrightarrow D_i$ and so on.  Since all
$R_i$ have the same length, as do all $C_i$ and $D_i$, this process is
guaranteed to terminate.

This is similar to, but \emph{not} the Knuth-Bendix algorithm: there
is no explicit concept of well-ordering, for example.  However, it
certainly contains many of the ``basic features'' of the algorithm
as described by Buchberger \citep{buchberger87}, and could be
considered, under restrictive conditions, as an embryonic version of
it.

\newpage
\section{Notes to accompany the translation}

\paragraph{Terminology:}
In the translation I have translated \emph{Zeichenreihen} literally as
``symbol sequence'', even though Post had already interpreted it
simply as ``string''.  I do this mainly to retain fidelity with Thue's
paper, where he on other occasions uses ``sequence'',
``sub-sequence'', and ``null sequence'' without using the word
\emph{Zeichen}.  

\subsection*{- Section I (pg 493)}

Here Thue briefly introduces the paper, noting that it follows from
his earlier work on trees \citep{thue10} and on sequences
that \emph{don't} contain overlapping sub-sequences \citep{thue12}.
These are the only two references in the paper even though e.g. the
prior work by Dehn was clearly relevant \citep{dehn11}.  This seems to
be a habit of Thue's: his 1912 paper only references
\citep{thue06}, and the 1910 and 1906 papers appear to have no
references at all.

With hindsight we can see Thue's work as fitting into the general format of
Hilbert's programme and the work of Emil Post (see e.g.\ \citep{demol13}), and it is interesting to note that Thue explicitly
understood his work as being of relevance to one of the ``most
fundamental problems''.  Yet Thue also limits \emph{his} programme quite
clearly: he will deal only with special cases of this problem.

\subsection*{- Section II  (pg 493)}

In this section Thue presents what Post termed a \emph{Thue system} as
a series of tuples of the form $(A_k, B_k)$.  Reading such a tuple as
a rule $A_k \leftrightarrow B_k$ allowing the replacement of a sub-string
$A_k$ with the string $B_k$ or vice versa, Thue defines the
concept of \emph{similar} sequences, and then \emph{equivalent} sequences
as the closure of this.  

Thue does not explicitly allow for empty sequences in the $A_i$ or $B_i$
(or anywhere else in the paper) and, in the absence of an identity
element, his \emph{Problem (I)} is thus the word problem for semi-groups.

The two special cases he deals with are:
\begin{itemize}
\item[(a)] Each $A_k$ and $B_k$ have the same length: thus, applying
  such a rule cannot change the length of the string, and there are
  only finitely many possibilities for permuting the symbols in these
  fixed-length strings.
\item[(b)] Each $A_k$  is longer than its corresponding $B_k$: so,
  applying a rule forwards will basically shrink the string, which can
  only be done a finite number of times.  Thue seems to slip into
  semi-Thue mode here, where he interprets the rules one-directionally
  as $A_k
  \rightarrow B_k$.  He also explicitly refers to this as a
  \emph{reduction}, and defines the term \emph{irreducible} (also
  defined in his 1910 paper).
\end{itemize}

Either of these restrictions give us a system that is
\emph{terminating}.  Thue proves by induction that if we also disallow
overlaps among the $A_k$ then the system must also be
\emph{confluent} (though he does not call it this), and thus the word
problem is decidable in this case.

\paragraph{Notation:} Thue distinguishes between:\\
\centerline{\begin{tabular}{rl}
$P \sim Q$ & $P$ can be transformed in 1 step into $Q$\\
$P = Q$ & $P$ can be transformed in 1 or more steps into $Q$\\
$P \equiv Q$ & $P$ is symbol-by-symbol identical to $Q$\\
\end{tabular}}

\subsection*{- Section III (pg 497)}

In this section Thue shifts the focus to systems based around some
given \emph{null sequence} $R$ which can be deleted from, or inserted
into, other strings.  This allows him to redefine the
terms \emph{similar} and \emph{equivalent} ``in respect of $R$'' in
this new context.  

In language theory terms, the null sequence here is not the empty
sequence $\epsilon$, but rather a \emph{nullable} sequence.  That is,
for the null sequence $R$, we implicitly have the rule $R
\leftrightarrow \epsilon$.

While it is not exactly explicit here, the
introduction of a null sequence brings us from semi-groups to monoids,
and Thue's \emph{Problem (II)} is the word problem in this case.

\paragraph{Overlaps}
The final few remarks of this section are of the utmost importance for
the rest of the paper.  Thue has already dealt (in \S II) with the
case where rules can't overlap, and he now addresses the case where
they \emph{can} overlap.  In the context of \S III, this means that we
have some string in which $R$ occurs twice as a sub-string, but in
overlapping configurations.  Calling this overlap $U$ we get:
$$
\begin{array}{r@{}c@{}l}
R \equiv C & U\\
           & U & D \equiv R\\
\end{array}
$$
where $CUD$ is a sub-string of the current string.
Thue derives the equivalence $C=D$ in respect of $R$ here, and will
make considerable use of this later.

\subsection*{- Section IV (pg 498)}
Having established the importance of identities of the type $CU \equiv
UD$, Thue investigates them further in this section.

First, Thue deals with a special case regarding power series (pp.
498-499).  If we have $CU \equiv UD$ then $C$ and $D$ must have the
same number of symbols.  If, in addition, they both have the same
number of symbols as $U$ then we must actually have $CD \equiv DC$.
Thue re-generalises this case slightly to consider situations where
$XY \equiv YX$ and $X$ and $Y$ are of different lengths.  In this case
Thue proves that both strings must be composed of some common factor
$\theta$, with $p$ copies in $X$, $q$ copies in $Y$ and thus $p+q$
copies in $XY$.

After dealing with a corollary ($AA$ containing $A$), Thue now returns
to the general case and sets up a kind of Euclidean algorithm for
factoring overlapping strings.  Starting with some string $U_0$, we factor this
into a quotient $C_1$ and remainder $U_1$, and then follow the same
process with the remainder.  

This process can be seen in action in the diagram on page 500.  Since
we know $S \equiv CU$ then $S$ must start with at least one $C$.  But
since $CU \equiv UD$, if $U$ has more symbols than $C$ then it must
also start with a $C$.  Hence $S \equiv CU$ must start with \emph{two}
$C$'s.  Repeating this process until what remains of $U$ becomes
shorter than $C$, we get $S \equiv C^n \alpha$: that is, $C$ divides
$n$ times into $S$ with remainder $\alpha$.  Following a similar
process with the $D$'s from the other end, we get the insight that
$C$, $D$ and $U$ must be formed from regular patterns of $\alpha$ and $\beta$.

An important point here is that we can factor $U$ as
$$U \equiv  (\alpha \beta)^{n-1} \alpha \equiv  \alpha (\beta
\alpha)^{n-1}$$
But this has the same format as the identity we began with;  i.e. it
has the form $U \equiv C_1U_1 \equiv U_1D_1$, and we can presumably
apply the same process, using $U$ this time instead of $S$.

Note that $U$ here is maximal and thus unique.  If we hadn't demanded that
$U$ be maximal, we could possibly have stopped dividing by $C$ at some
earlier stage $m<n$ and then get a larger overlap
$$U \equiv  (\alpha \beta)^{n-m} \alpha \equiv  \alpha (\beta
\alpha)^{n-m}$$
where $C \equiv (\alpha \beta)^m$ and $D \equiv (\beta\alpha)^m$.

In the final result of this section (pg 502) Thue shows that this process can
work `backwards'.  Just as we can start with some $M$ and derive $N$
as the largest overlap with remainder $X$ and $Y$, we can in a similar
manner derive a string $T$ for which $M$ is the largest overlap with
remainder $X$ and $Y$.

\subsection*{- Section V (pg 503)}

In this section Thue considers the relationship between a presentation
in terms of some null sequence $R$ (as in \S III), and one in terms of
a set of equations (as in \S II).  In particular he wants to know under
what circumstances a set of equations (such as (1)) can
\emph{adequately} represent the null sequence.  In algebraic terms,
one might regard this as asking the question: when when can a set of
equations in the presentation of a semi-group adequately model the
presentation of a monoid (which could include equations of the form $R
= 1$).

Two sequences that are provably equivalent according to the semi-group
equations are called ``parallel'' and Thue uses the notation $P \jpPar
Q$.  This notation is only used in this section and in examples 1 and 5
of \S VII.  In general, if $R$ is the identity in a monoid, then we
must have for any other element $z$ that $zR = z = Rz$.  For semi-group
equations like (1) to model this, we must have the power to prove all
equations of this kind; Thue splits this into two parts: the equations are
\begin{itemize}
\item \emph{complete} (vollst\"andiges) if we can prove $zR \jpPar Rz$
\item \emph{perfect} (vollkommenes) if we can prove $RA \jpPar RB$
implies $A \jpPar B$ 
\end{itemize}

Thus we have an algorithm for dealing with sequences containing the
null sequence.  The theorem on page 504 sets this up, and the
algorithm is presented on pg 505.
Given some sequence $P$, a complete system allows us
to ``move around'' any occurrence of the
null sequence $R$ in $P$, thus deriving all other
sequences of the same length that differ only in the position of $R$.
A perfect system then allows us to delete null sequences on the left,
forming a new  collection of (shorter) parallel sequences.  Repeating
this process we eventually get to a set of equivalent irreducible sequences.
Given some other sequence $Q$, it is equivalent to $P$ if it reduces
down to the same set of irreducible sequences.

\subsection*{- Section VI (pg 506)}

In this section Thue imposes some fairly severe restrictions on the
format of the equations and shows that this helps determining the
matching between sequences.  He reintroduces the terminology from \S
III (no null sequences in \S VI), and then in the main theorem on page 506
he restricts the format of allowable equations based on the first
symbol on each side.  Thus given some sequence $P$, if I apply a
series of the equations to $P$, then I can guarantee that (at worst)
each two applications will fix a leftmost symbol in $P$ for the rest
of the derivation.

For example, if I apply a rule of the form $A_i \rightarrow B_i$,
then, if the leftmost symbol of $P$ changes, it can only change from $x_i$ to
$y_i$.  But since no other $A_j$ or $B_j$ starts with $y_i$, the
leftmost symbol is effectively fixed for the rest of the derivation.
(I could choose $B_i \rightarrow A_i$, but this just makes the first
step redundant).

Alternatively, if I could apply a rule of the form $B_i \rightarrow
A_i$ and change the leftmost symbol of $P$, it must change from $y_i$
to $x_i$.  But now if I wish to change the leftmost symbol again I
must this time pick some rule of the form $A_i \rightarrow B_i$, and
the argument from the previous paragraph then holds.

Thue presents this from a different perspective: if the leftmost part
of two equivalent strings are equivalent, then so are the remaining
rightmost parts.  This is proved methodically on pages 506-509.

\subsection*{- Section VII (pg 510)}

In this section Thue gives 5 examples of systems of equations that are
complete and perfect.

The first example is a set of equations derived from a factoring of
the null sequence $R$ using the method of \S IV.  Two points worth
noting here: 
\begin{itemize}
\item just below the identities in (6) we are told that all the
$Y_i$ and $X_r$ are different and 
\item just below the equivalences in (7) it is noted that $X_1$ begins
  with $X_r$ (and, thus, so do all the left-hand-sides of the equations)
\end{itemize}
Taken together, this means that the equivalences in (7) have the
format required for the main theorem in \S VI.  In particular,
applying this theorem with $R$ in place of $C$ and $D$ lets us
conclude that whenever $RM = RN$ then $M=N$, meaning that (7) forms a
perfect set of equivalences.

This approach is used (implicitly) to prove that the equation systems
are perfect in examples 1, 2, 3 and 4.  Note in examples 2, 3 and 4
that $R$ is chosen so that it can be broken down into the usual
``overlap'' pattern
$$
\begin{array}{r@{}c@{}l}
R \equiv C & U\\
           & U & D \equiv R\\
\end{array}
$$
which then yields equivalences of the form $C=D$.

Example 5 is a little different, since the equations are not
constructed to be obviously perfect following the template of the
others.  Actually \emph{proving} that the system is perfect requires
considerable effort, stretching from page 514-516.

\subsection*{- Section VIII (pg 516)}

Thue's ``few remarks'' here amount to an algorithm for constructing a
system of equivalences starting from a given null sequence $R$.

Given some null sequence $R$, Thue shows how to generate an initial
set of equations based on any overlaps (as in example 2 of the
previous section).  That is, we form the equation $C=D$ for each
possible overlap $U$ that satisfies $R \equiv CU \equiv UD$.  This
implies that $C$ and $D$ must have the same length, and they must have
fewer symbols than $R$.  Having derived a set of such equations, we
can then apply them to $R$ to derive another set of null sequences
equivalent to $R$: these all have the same length as the original $R$.
We can continue in this way, alternating between deriving new set of null
sequences $S_\theta$ and new sets of equations $E_\theta$.
Since the length of all the $P$s, $Q$s and $R$s are bounded, the process
must terminate, as noted on page 519.

In the following discussion (pp. 519-521) he shows how to use such a
system $\delta$ and to `minimise' these equations to a derived system
$\epsilon$.  He then proves a series of theorems demonstrating the
`minimality' of this system.

In the final theorem starting on page 522 it may not be obvious that
the eight cases listed exhaust all the possibilities for the configuration
of the overlap between $M \equiv aR_zb$ and $N \equiv c R_\mu d$.  The
key here is to work out the overlap between $M$ and $N$ and to note
that it (mostly) shrinks as we move through the cases.

Since both $M$ and $N$ are divided into three sub-strings, we can
characterise the overlap by categorising the degree to which each
sub-string is involved in it.  For example, the maximal amount of overlap
(assuming the strings aren't identical) would be given by the
following configuration

$$\begin{array}{@{\extracolsep{-8pt}}*{7}{c}}
\jpLetter{2}{a} & \jpLetter{2}{R_z} & \jpLetter{2}{b} \\
\jpOverBrace{2} & \jpOverBrace{2} & \jpOverBrace{2} \\
\jpLetter{1}{C} &\jpLetter{1}{e} &\jpLetter{1}{f} 
&\jpLetter{1}{g} &\jpLetter{1}{h} &\jpLetter{1}{i} &\jpLetter{1}{D} \\
\jpBlank{1} & \jpUnderBrace{2} & \jpUnderBrace{2} & \jpUnderBrace{2} \\
\jpBlank{1} &\jpLetter{2}{c} & \jpLetter{2}{R_\mu} & \jpLetter{2}{d} \\
\end{array}$$

In this configuration, the overlap $U \equiv efghi$
involves \emph{part} of $a$ and $d$ and \emph{all} 
of the other four sub-strings.   We can get the remaining configurations by
sliding $M$ to the left (or, equivalently, $N$ to the right).
Characterising the involvement of the six sub-strings as $P$, $A$ or $N$
for \emph{part}, \emph{all} or \emph{none} respectively, we can
actually track nine cases as we decrease the overlap.  I found it useful
to enumerate the first eight of these as follows:

$$
\begin{array}{c|c|c|c}
& & {M \equiv} & {N \equiv} \\
& U \equiv & a R_z b & c R_\mu d\\
\hline
1 &efghi & PAA & AAP\\
2 & efgb & PAA & APN \\
3 & cefgb & PAA & AAP \\
4 & cfb & NPA & APN \\
5 & fgh & NPA & APN \\
6 & fb & NPA & PNN\\
7 & cf & NNP & APN \\
8 & f & NNP & PNN \\ 
\end{array}
$$

The ninth case,
which we could characterise as $NNP$ versus $AAP$ is actually
impossible, since it would result in $R_\mu$ being a sub-string of
$R_z$, and all the strings $R$ are supposed to have the same length.

The eight cases Thue deals with are illustrated in Figure \ref{fig:eightCases}.

\begin{figure}
$$
\begin{array}{|rc|rc|}
\hline
&&&\\
1. & U \equiv efghi & 2. & U \equiv efgb,\; D \equiv hd \\
&&&\\
& \begin{array}{@{\extracolsep{-8pt}}*{7}{c}}
\jpLetter{2}{a} & \jpLetter{2}{R_z} & \jpLetter{2}{b} \\
\jpOverBrace{2} & \jpOverBrace{2} & \jpOverBrace{2} \\
\jpLetter{1}{C} &\jpLetter{1}{e} &\jpLetter{1}{f} 
&\jpLetter{1}{g} &\jpLetter{1}{h} &\jpLetter{1}{i} &\jpLetter{1}{D} \\
\jpBlank{1} & \jpUnderBrace{2} & \jpUnderBrace{2} & \jpUnderBrace{2} \\
\jpBlank{1} &\jpLetter{2}{c} & \jpLetter{2}{R_\mu} & \jpLetter{2}{d} \\
\end{array}
&
& \begin{array}{@{\extracolsep{-8pt}}*{7}{c}}
\jpLetter{2}{a} & \jpLetter{2}{R_z} & \jpLetter{1}{b} \\
\jpOverBrace{2} & \jpOverBrace{2} & \jpOverBrace{1} \\
\jpLetter{1}{C} &\jpLetter{1}{e} &\jpLetter{1}{f} 
&\jpLetter{1}{g} & \jpLetter{1}{b} &\jpLetter{1}{h} &\jpLetter{1}{d} \\
\jpBlank{1} & \jpUnderBrace{2} & \jpUnderBrace{3} & \jpUnderBrace{1} \\
\jpBlank{1} &\jpLetter{2}{c} & \jpLetter{3}{R_\mu} & \jpLetter{1}{d} \\
\end{array}
\\
&&&\\ \hline
&&&\\
3. & U \equiv cefgb & 4. & C \equiv ae,\; U \equiv cfb,\; D \equiv gd  \\
&&&\\
& \begin{array}{@{\extracolsep{-8pt}}*{7}{c}}
\jpLetter{3}{a} & \jpLetter{2}{R_z} & \jpLetter{1}{b} \\
\jpOverBrace{3} & \jpOverBrace{2} & \jpOverBrace{1} \\
\jpLetter{1}{C} &\jpLetter{1}{c} &\jpLetter{1}{e} 
&\jpLetter{1}{f} &\jpLetter{1}{g} &\jpLetter{1}{b} &\jpLetter{1}{D} \\
\jpBlank{1} & \jpUnderBrace{1} & \jpUnderBrace{2} & \jpUnderBrace{3} \\
\jpBlank{1} &\jpLetter{1}{c} & \jpLetter{2}{R_\mu} & \jpLetter{3}{d} \\
\end{array}
&
& \begin{array}{@{\extracolsep{-8pt}}*{7}{c}}
\jpLetter{1}{a} & \jpLetter{3}{R_z} & \jpLetter{1}{b} \\
\jpOverBrace{1} & \jpOverBrace{3} & \jpOverBrace{1} \\
\jpLetter{1}{a} &\jpLetter{1}{e} &\jpLetter{1}{c} 
&\jpLetter{1}{f} &\jpLetter{1}{b} &\jpLetter{1}{g} &\jpLetter{1}{d} \\
\jpBlank{2} & \jpUnderBrace{1} & \jpUnderBrace{3} & \jpUnderBrace{1} \\
\jpBlank{2} &\jpLetter{1}{c} & \jpLetter{3}{R_\mu} & \jpLetter{1}{d} \\
\end{array}
\\
&&&\\ \hline
&&&\\
5. & C \equiv ae,\; U \equiv fgh,\; D \equiv id   & 
6. & C \equiv ae,\; U \equiv fb,\; D \equiv gR_\mu d   \\
&&&\\
& \begin{array}{@{\extracolsep{-8pt}}*{7}{c}}
\jpLetter{1}{a} & \jpLetter{2}{R_z} & \jpLetter{2}{b} \\
\jpOverBrace{1} & \jpOverBrace{2} & \jpOverBrace{2} \\
\jpLetter{1}{a} &\jpLetter{1}{e} &\jpLetter{1}{f} 
&\jpLetter{1}{g} &\jpLetter{1}{h} &\jpLetter{1}{i} &\jpLetter{1}{d} \\
\jpBlank{2} & \jpUnderBrace{2} & \jpUnderBrace{2} & \jpUnderBrace{1} \\
\jpBlank{2} &\jpLetter{2}{c} & \jpLetter{2}{R_\mu} & \jpLetter{1}{d} \\
\end{array}
&
& \begin{array}{@{\extracolsep{-8pt}}*{7}{c}}
\jpLetter{1}{a} & \jpLetter{2}{R_z} & \jpLetter{1}{b} \\
\jpOverBrace{1} & \jpOverBrace{2} & \jpOverBrace{1} \\
\jpLetter{1}{a} &\jpLetter{1}{e} &\jpLetter{1}{f} 
&\jpLetter{1}{b} &\jpLetter{1}{g} &\jpLetter{1}{R_\mu} &\jpLetter{1}{d} \\
\jpBlank{2} & \jpUnderBrace{3} & \jpUnderBrace{1} & \jpUnderBrace{1} \\
\jpBlank{2} &\jpLetter{3}{c} & \jpLetter{1}{R_\mu} & \jpLetter{1}{d} \\
\end{array}
\\
&&&\\ \hline
&&&\\
7. & C \equiv a R_z e,\; U \equiv cf,\; D \equiv g d    & 
8. & C \equiv a R_z e,\; U \equiv f,\; D \equiv gR_\mu d  \\
&&&\\
& \begin{array}{@{\extracolsep{-8pt}}*{7}{c}}
\jpLetter{1}{a} & \jpLetter{1}{R_z} & \jpLetter{3}{b} \\
\jpOverBrace{1} & \jpOverBrace{1} & \jpOverBrace{3} \\
\jpLetter{1}{a} &\jpLetter{1}{R_z} &\jpLetter{1}{e} 
&\jpLetter{1}{c} &\jpLetter{1}{f} &\jpLetter{1}{g} &\jpLetter{1}{d} \\
\jpBlank{3} & \jpUnderBrace{1} & \jpUnderBrace{2} & \jpUnderBrace{1} \\
\jpBlank{3} &\jpLetter{1}{c} & \jpLetter{2}{R_\mu} & \jpLetter{1}{d} \\
\end{array}
&
& \begin{array}{@{\extracolsep{-8pt}}*{7}{c}}
\jpLetter{1}{a} & \jpLetter{1}{R_z} & \jpLetter{2}{b} \\
\jpOverBrace{1} & \jpOverBrace{1} & \jpOverBrace{2} \\
\jpLetter{1}{a} &\jpLetter{1}{R_z} &\jpLetter{1}{e} 
&\jpLetter{1}{f} &\jpLetter{1}{g} &\jpLetter{1}{R_\mu} &\jpLetter{1}{d} \\
\jpBlank{3} & \jpUnderBrace{2} & \jpUnderBrace{1} & \jpUnderBrace{1} \\
\jpBlank{3} &\jpLetter{2}{c} & \jpLetter{1}{R_\mu} & \jpLetter{1}{d} \\
\end{array}
\\
&&&\\ \hline
\end{array}
$$
\caption{The eight cases of the final theorem in \S VIII (pages 522-524)}
\label{fig:eightCases}
\end{figure}

\newpage

\bibliographystyle{plainnat}
\bibliography{thue,thue-related}

\begin{thebibliography}{18}
\providecommand{\natexlab}[1]{#1}
\providecommand{\url}[1]{\texttt{#1}}
\expandafter\ifx\csname urlstyle\endcsname\relax
  \providecommand{\doi}[1]{doi: #1}\else
  \providecommand{\doi}{doi: \begingroup \urlstyle{rm}\Url}\fi

\bibitem[Berstel(1995)]{berstel95}
Jean Berstel.
\newblock {A}xel {T}hue's papers on repetitions in words: a translation.
\newblock Publications du {LaCIM}, Universit\'e du Qu\'ebec \`a Montr\'eal,
  1995.

\bibitem[Buchberger(1987)]{buchberger87}
Bruno Buchberger.
\newblock History and basic features of the critical-pair/completion procedure.
\newblock \emph{Journal of Symbolic Computation}, 3\penalty0 (1-2):\penalty0
  3--38, 1987.

\bibitem[Chomsky(1959)]{chomsky59}
Noam Chomsky.
\newblock On certain formal properties of grammars.
\newblock \emph{Information and Control}, 2\penalty0 (2):\penalty0 137--167,
  June 1959.

\bibitem[Church(1936)]{church36}
Alonzo Church.
\newblock A bibliography of symbolic logic.
\newblock \emph{The Journal of Symbolic Logic}, 1\penalty0 (4):\penalty0 pp.
  121--216, 1936.

\bibitem[Dehn(1911)]{dehn11}
M.~Dehn.
\newblock \"{U}ber unendliche diskontinuierliche gruppen.
\newblock \emph{Mathematische Annalen}, 71\penalty0 (1):\penalty0 116--144,
  1911.

\bibitem[{DeMol}(2013)]{demol13}
Liesbeth {DeMol}.
\newblock Formalism. {T}he success(es) of a failure.
\newblock In A.~Moretti A.~Moktefi and F.~Schang, editors, \emph{Let's be
  logical}. College publications, 2013.
\newblock (to appear).

\bibitem[Hedlund(1967)]{hedlund67}
G.A. Hedlund.
\newblock Remarks on the work of {A}xel {T}hue on sequences.
\newblock \emph{Nordisk Matematisk Tidskrift}, 15:\penalty0 148--150, 1967.

\bibitem[Hilbert(1902)]{hilbert02}
David Hilbert.
\newblock Mathematical problems.
\newblock \emph{Bulletin of the American Mathematical Society}, 8\penalty0
  (10):\penalty0 437--479, 1902.

\bibitem[Nagell et~al.(1977)Nagell, Selberg, Selberg, and Thalberg]{nagell77}
Trygve Nagell, Atle Selberg, Sigmund Selberg, and Knut Thalberg, editors.
\newblock \emph{Selected Mathematical Papers of Axel Thue}.
\newblock Universitetsforlaget, Oslo, 1977.

\bibitem[Post(1943)]{post43}
Emil~L. Post.
\newblock Formal reductions of the general combinatorial decision problem.
\newblock \emph{American Journal of Mathematics}, 65\penalty0 (2):\penalty0
  197--215, April 1943.

\bibitem[Post(1946)]{post46}
Emil~L. Post.
\newblock A variant of a recursively unsolvable problem.
\newblock \emph{Bulletin of the American Mathematical Society}, 52\penalty0
  (4):\penalty0 264--268, 1946.

\bibitem[Post(1947)]{post47}
Emil~L. Post.
\newblock Recursive unsolvability of a problem of {T}hue.
\newblock \emph{Journal of Symbolic Logic}, 12\penalty0 (1):\penalty0 1--11,
  March 1947.

\bibitem[Scholz and Pullum(2007)]{scholz07}
Barbara~C. Scholz and Geoffrey~K. Pullum.
\newblock Tracking the origins of transformational generative grammar.
\newblock \emph{Journal of Linguistics}, 43\penalty0 (3):\penalty0 pp.
  701--723, 2007.

\bibitem[Steinby and Thomas(2000)]{steinby00}
M.~Steinby and W.~Thomas.
\newblock Trees and term rewriting in 1910: On a paper by {A}xel {T}hue.
\newblock \emph{EATCS Bull.}, 72:\penalty0 256--269, 2000.

\bibitem[Thue(1906)]{thue06}
Axel Thue.
\newblock {\"Uber unendliche Zeichenreihen}.
\newblock \emph{Christiana Videnskabs-Selskabs Skrifter, I. Math.-naturv.
  Klasse}, 7, 1906.

\bibitem[Thue(1910)]{thue10}
Axel Thue.
\newblock {Die L\"osung eines Spezialfalles eines generellen logischen
  Problems}.
\newblock \emph{Christiana Videnskabs-Selskabs Skrifter, I. Math.-naturv.
  Klasse}, 8, 1910.

\bibitem[Thue(1912)]{thue12}
Axel Thue.
\newblock {\"Uber die gegenseitige Lage gleicher Teile gewisser Zeichenreihen}.
\newblock \emph{Christiana Videnskabs-Selskabs Skrifter, I. Math.-naturv.
  Klasse}, 1, 1912.

\bibitem[Thue(1914)]{thue14}
Axel Thue.
\newblock {Probleme \"uber Ver\"anderungen von Zeichenreihen nach gegebenen
  Regeln}.
\newblock \emph{Christiana Videnskabs-Selskabs Skrifter, I. Math.-naturv.
  Klasse}, 10, 1914.

\end{thebibliography}

\newpage


\titleformat{\section}%
  [hang]
  {\normalfont\huge}
  {\thesection}
  {1em}
  {\filcenter\S\;}
  []

\titleformat{\subsection}%
  [hang]
  {\Large\sffamily}
  {\thesubsection}
  {1em}
  {\filcenter}
  []

\newcommand\UseKEquationNumbering{
  \renewcommand{\theequation}{K}
}

\newcommand\UseGreekEquationNumbering{
  \renewcommand{\theequation}{\greek{equation}}
  \setcounter{equation}{0}
}

\newtheorem{problem}{Problem}
\renewcommand{\theproblem}{(\Roman{problem})}

\setcounter{page}{493}
\newcommand\mypage[1]{\newpage}



\noindent
{\Large\em Problems concerning the transformation of
symbol sequences according to given rules}\\

\medskip
\noindent
Axel Thue\marginnote{This translation by James F. Power <jpower@cs.nuim.ie>}

\medskip
\noindent
1914\marginnote{\today}

\begin{abstract}
\noindent This document is a translation of Axel Thue's paper 
\marginnote{Based on the version published in his
  \emph{Selected Mathematical Papers} as paper \#28, pp. 493-524.
  Page numbers in this document follow that pagination.}
  \emph{Probleme
  \"uber Ver\"anderungen von Zeichenreihen nach gegebenen Reglen}
(Kra. Videnskabs-Selskabets
Skrifter. I. Mat. Nat.Kl. 1914. No. 10)
\marginnote{Any margin notes (like this) are not part of the original
  paper, but typos noted here are typos in the original paper.}
\end{abstract}

\section{I}\label{sec:one} 

In a previous work${}^1$ I have posed the general
question whether two given concepts depicted as trees, but defined in
different ways, must be equivalent to each other.

In this paper I will deal with a problem concerning the transformation
of symbol sequences using rules.  This problem, that in certain respects is a
special case of one of the most fundamental problems that can be
posed, is also of immediate significance for the general case.
Since this task seems to be extensive and of the utmost difficulty,
I must be satisfied with only treating the question in a piecewise and
fragmentary manner.

In a previous year's work${}^2$ I have already solved a special case
concerning symbol sequences.  On this occasion I will just settle some
simple cases of the aforementioned general problem.  I will not enter
into a discussion here on the wider significance of investigations of
this type.

\section{II}

We are given two series of symbol sequences:
$$\begin{array}{ccccc}
A_1,& A_2,& A_3,& \ldots,& A_n\\
B_1,& B_2,& B_3,& \ldots,& B_n,
\end{array}$$
where each symbol in each sequence $A$ and in each sequence $B$ is a
symbol from some group of given symbols.

\vfill

\rule{.3\textwidth}{.5pt}\\[1ex]
{\footnotesize\begin{tabular}{lp{.8\textwidth}}
${}^1$ &Die L\"osung eines
  Spezialfalles eines generellen logischen Problems.  (Christiana
  Videnskabsselskabs Skrifter, 1910.)
\end{tabular}}\marginnote{\emph{Selected Mathematical
    Papers} pp. 273-310}\\
{\footnotesize\begin{tabular}{lp{.8\textwidth}}
${}^2$ & \"Uber die
  gegenseitige Lage gleicher Teile gewisser Zeichenreihen.  (Christiana
  Videnskabsselskabs Skrifter, 1912.)
\end{tabular}}\marginnote{\emph{Selected Mathematical
    Papers} pp. 413-478}\\

\mypage{494}

For each value of $k$ we will call $A_k$ and $B_k$ \emph{corresponding
 sequences}.

If, given two arbitrary sequences $P$ and $Q$, one can get one from
the other by replacing some subsequence $A$ or $B$ by its
corresponding sequence, then we say that $P$ and $Q$ are called
\emph{similar sequences} with reference to the corresponding sequences
$A$ and $B$.  We indicate this by writing 
$$P \sim Q$$

The sequences $\alpha A_h \beta$ and $\alpha B_h \beta$, where
$\alpha$ and $\beta$ denote symbol sequences, are thus (for example) similar
sequences.

If any two two symbol sequences $X$ and $Y$ are procured in this way, then one
can find a series of symbol sequences
$$C_1, C_2, \ldots, C_r$$
such that $X$ and $C_1$, then $C_r$ and $Y$, and finally $C_h$ and
$C_{h+1}$ for each $h$ are equivalent sequences, so then we have thus:
$$X \sim C_1 \sim C_2 \sim \ldots \sim C_r \sim Y,$$
then we say that $X$ and $Y$ are \emph{equivalent sequences} in respect of
the given sequences $A$ and $B$.

We denote this by means of the equation
$$X = Y$$

When $P \sim Q$ we also have $P = Q$.  Further, we have $A_k \sim B_k$
and $A_k = B_k$.

\bigskip

We can now pose the major general question \ref{prob:I}:

\begin{problem}
For any arbitrary given sequences $A$ and $B$, to find a method, where
one can always calculate in a predictable number of operations,
whether or not two arbitrary given symbol sequences are equivalent in
respect of the sequences $A$ and $B$.
\label{prob:I}
\end{problem}

This problem is easily solved in the following two cases (a) and (b).

\begin{enumerate}
\item[(a)] $A_k$ and $B_k$ contain equal number of symbols for each value of
$k$.
\mypage{495}

Here, either two sequences $X$ and $Y$ are equivalent, or one can find
sequences $C_1, C_2, \ldots, C_r$ such that:
$$C_0 \sim C_1 \sim C_2 \sim \ldots \sim C_r \sim C_{r+1},$$
where $C_0$ and $C_{r+1}$ denote $X$ and $Y$ respectively, then any
two of the sequences of $C$ have equal number of symbols, but can be
assumed to be different from each other.

$r$ must consequently fall under a predictable limit, and the problem
is thus solved.

\item[(b)] $A_k$ contains more symbols than $B_k$ for each value of
  $k$. \marginnote{This seems to assume we're operating semi-Thue system $A_k
  \rightarrow B_k$}
The sequence $A$ is in addition so constituted that any two arbitrary
subsequences $A_p$ and $A_q$ must always lie completely outside each
other for any values of $p$ and $q$.

We assume, in other words, that no sequence in $A$ can be a subsequence
of another sequence in $A$, while further two arbitrary possible
subsequences $A_p$ and $A_q$ are not allowed to have any common part.
\end{enumerate}

We use the term \emph{irreducible sequence} to refer to any sequence which
contains no subsequence $A$.

Case (b) of the aforementioned problem can now be solved as follows:

Through repeated reductions of an arbitrary given symbol sequence $S$ we can
only get a single irreducible sequence.
\marginnote{That is, the system is confluent.}
Here, in each reduction a
subsequence $A$ of the given sequence (or any sequence obtained from
it via an previous reduction) is replaced by its corresponding
sequence $B$.

This \marginnote{The proof proceeds by induction over the number of
  symbols in $S$.}
statement must be correct in the case where $S$ contains only a
single symbol.

It is also immediately apparent that if the statement is correct when
the number of symbols in $S$ is less than some number $t$, then it
must also be correct when the number of symbols in $S$ is equal to
$t$.

In particular if $S$ is irreducible, the case is immediately clear.  It
also follows when $S$ only contains a single subsequence $A$, i.e.\
$$S \equiv M A_k N,$$
where we use the symbol $\equiv$ to denote the identity.  $S$ can then
only be reduced to the same irreducible sequence as $M B_k
N$.\marginnote{Since we know that $A_k$ has more symbols than $B_k$, the
  inductive hypothesis applies to prove this case.}

Finally we suppose that $S$ is gradually reduced, in two different ways,
to the two irreducible sequences $P$ and $Q$ respectively.  

\mypage{496}

In the
first reduction $S$ is reduced by a single reduction to $H$, and then
by a sequence of reductions to $P$.  In the
second reduction $S$ is reduced by a single reduction to $K$, and then
by a sequence of reductions to $Q$.

The case is immediately clear when the two first reductions are the
same as each other, i.e.\ $H \equiv K$.  If this is not the case, we
can write:
$$\begin{array}{rcl}
S &\equiv& M A_p L A_q N\\
H &\equiv& M B_p L A_q N\\
K &\equiv& M A_p L B_q N\\
\end{array}$$
where one or more of the sequences $M$, $L$ and $N$ are allowed to be absent.

However $H$ and $K$ are so constituted that they contain fewer symbols than
$t$, and can then be reduced to a single irreducible sequence $M B_p L
B_q N$.  That is, 
$$P \equiv Q$$

Since any two similar rows, and even two equivalent rows, can be
reduced in this way only to a single irreducible sequence, this proves
the result.

\bigskip

\begin{quotation}
\marginnote{The next few paragraphs are in a reduced font in Thue's
  paper, thus presumably a kind of sidenote.}
Instead of problem \ref{prob:I}, one can set up the still more general
question:

Suppose $P$ and $Q$ signify two arbitrary symbol sequences, and that
each symbol that occurs in them is different from those in the series
$A$ and $B$.  Then, find a general method by which it is possible to
decide whether any of the symbols of $P$ and $Q$ can be replaced by
such symbol sequences, so that the symbol sequences $P'$ and $Q'$
obtained from $P$ and $Q$ in this way are equivalent.

We assume that symbols that are equal to one another are only replaced
by sequences that are equal to one another.

\medskip

We can also generalise problem \ref{prob:I} in another way.

Given two arbitrary symbol sequences $P$ and $Q$, one can 
get one of them from the other by replacing a subsequence $A'$ with
another sequence $B'$, where $A'$ and $B'$ have been obtained in such
a way that one can write sequences in place of the symbols of two
corresponding sequences $A$ and $B$, such that $A$ and $B$ in this way
turn to $A'$ and $B'$ respectively, then we can - with a new meaning
of the words - define $P$ and $Q$ to be similar.

In a corresponding manner we can define equivalent sequences and pose
the question: how can one always decide whether two sequences are
equivalent, or whether in place of the sequences one can write two
such sequences that in this way the sequences become equivalent.

\end{quotation}

We wish now to deal with an important special case of problem \ref{prob:I}.

We wish to give a new definition of the concepts \emph{similarity} and
\emph{equivalence}.

\mypage{497}

\section{III}

Let $R$ signify an arbitrary given symbol sequence.  Two arbitrary
sequences are said to be \emph{sequences similar to each other with
  respect to $R$} when we can get one from the other by removing
a subsequence $R$.

The sequences 
$$MRN \hbox{ and } MN,$$
where one of the arbitrary sequences $M$ and $N$ can of course be
missing, are thus examples of similar sequences.

When $P$ and $Q$ are similar sequences, we can indicate this by
writing
$$P \sim Q.$$

If we are given two arbitrary sequences $P$ and $Q$ such that
sequences $C_1, C_2, \ldots, C_r$ exist, where $X$ and $C_1$, also
$C_r$ and $Y$, and finally $C_h$ and $C_{h+1}$ for each value of $k$,
are similar sequences, so that
$$X \sim C_1 \sim C_2 \sim \cdots \sim C_r \sim Y$$
then we will call $X$ and $Y$ \emph{equivalent sequences in respect of
  $R$}.  We indicate this by writing
$$X = Y$$

Equivalent sequences can always be transformed into one another by
removal and insertion of sequence $R$.

If we have $P \sim Q$ then we also have $P = Q$.

If we have $A=C$ and $B=C$ then we also have $A=B$.

If $X$ and $Y$ are two arbitrary equivalent sequences, then one can
find such sequences $H$ and $K$ that
\begin{gather*}
H_0 \sim H_1 \sim H_2 \sim \cdots \sim H_p,\\
K_0 \sim K_1 \sim K_2 \sim \cdots \sim K_q,
\end{gather*}
where $H_0 \equiv X$, $K_0 \equiv Y$ and $H_p \equiv K_q$,\mypage{497}
and one can always get from $H_{r-1}$ to $H_r$ and $K_{s-1}$ to $K$ by
removing a null sequence \marginnote{``Nullreihe'' is first used here, but
  then defined at the end of this section} $R$.

If we have
$$X \sim C_1 \sim C_2 \sim \cdots \sim C_m \sim Y$$
where e.g.
$$\begin{array}{rcl}
C_{t-1} &\equiv& xRyz\\
C_{t} &\equiv& xyz\\
C_{t+1} &\equiv& xyRz\\
\end{array}$$
then we also have
$$\cdots \sim C_{t-1} \sim xRyRz \sim C_{t+1} \sim \cdots$$
etc.

\bigskip

One can now state the major task \ref{prob:II}:

\begin{problem}
Given an arbitrary sequence $R$, to find a method where one can
always decide in a finite number of investigations whether or not two
arbitrary given sequences are equivalent with respect to $R$.
\label{prob:II}
\end{problem}

The ultimate goal of our discussion now lies in giving the solutions
for some examples of this task.

We have shown earlier that the case is clear when two subsequences of
$R$ can never have common part.

The difficulty arises when the opposite case occurs.  If two
\marginnote{This pattern for overlapping sequences, $CU \equiv UD$,
is a recurring
  motif in the remainder of the paper}
subsequences of $R$ can have a common part $U$ then we can write
$$R \equiv CU \equiv UD$$
or
$$C \sim C(UD) \equiv (CU)D \sim D$$
or 
$$C = D$$

The sequence $R$ is called a \emph{null sequence}.

\section{IV}

By $T^n$, where $T$ denotes and arbitrary symbol sequence, we wish to
signify the construction of a sequence 
$$TT \cdots T$$
from $n$ copies of the sequence $T$.

We say that $T^n$ is called a \emph{Power series}.

\medskip

If $X$ and $Y$ are two sequences so constituted that
$$XY \equiv YX$$ \mypage{499}
then there exists a sequence $\theta$ such that
$$X \equiv \theta^p, \qquad Y \equiv \theta^q$$

The case where $X$ and $Y$ contain equally many symbols is clearly
true.  For then
$$X \equiv Y$$
or 
\begin{gather*}
\theta \equiv X \equiv Y\\
p = q = 1\\
\end{gather*}

The case where $XY$ contains only two symbols is thus clearly true.

However, if the case is true when $XY$ has fewer than $m$ symbols,
then it must also be true when $XY$ has exactly $m$ symbols.

Suppose here that, for example, $X$ is composed of more symbols than
$Y$; then one has:
$$X \equiv YZ$$
or
$$(YZ)Y \equiv XY \equiv YX \equiv Y(YZ)$$
or
$$YZ \equiv ZY$$
or
$$Z \equiv \theta^\gamma, \quad Y \equiv \theta^\delta, \quad X
\equiv \theta^{\gamma+\delta}$$

\bigskip

If a sequence $AA$ is composed of an inner subsequence $A$, we can
then write:

$$\renewcommand\arraystretch{.4}
\begin{array}{@{\extracolsep{-8pt}}*{8}{c}}
\jpLetter{4}{A} & \jpLetter{4}{A} \\
\jpOverBrace{4} & \jpOverBrace{4}\\
\hline
\jpLetterR{1}{x} & \jpLetterR{3}{y} & \jpLetterR{1}{x} \\
\cline{2-5}
\jpBlank{1} & \jpUnderBrace{4}\\[3pt]
\jpBlank{1} & \jpLetter{4}{A}
\end{array}$$

$$A \equiv xy \equiv yx$$
or consequently
$$x \equiv \theta^p, \quad y \equiv \theta^q, 
\quad A \equiv \theta^{p+q}.$$

\bigskip

Let $U_0$ now denote an arbitrary given symbol sequence.  
\marginnote{This gives us a kind of Euclidean (GCD) algorithm for
  sequences, with each $U_i$ as the quotient and $C_i$ as the remainder.}
We can then
define such sequences $U$, $C$ and $D$ such that:
$$\begin{array}{ccc}
U_0 &\equiv& C_1 U_1 \equiv U_1D_1\\
U_1 &\equiv& C_2 U_2 \equiv U_2D_2\\
\hdotsfor{3}\\
U_{r-1} &\equiv& C_r U_r \equiv U_rD_r\\
\end{array}$$
where $U_p$ for any value of $p$ is the largest sequence for which one
can find such sequences $C_p$ and $D_p$ that
$$U_{p-1}  \equiv  C_pU_p \equiv U_pD_p,$$
\mypage{500}
where two subsequences $U_r$ in the sequence can never have a common part.

If one has
$$U_0  \equiv  CU \equiv UD,$$
then $U$ must be equal to one of the sequences $U_1, U_2, \ldots,
U_r$, as can be immediately seen.

If $S$ denotes an arbitrary given symbol sequence and $U$ the largest
sequence for which one can find two sequences $C$ and $D$ such that
$$S \equiv CU \equiv UD$$
or
$$CS \equiv CUD \equiv SD$$
then first
$$S \equiv CC \cdots C \alpha \equiv C^n \alpha,$$

where the sequence $\alpha$ is either wholly missing, or is composed
of fewer symbols than $C$.

Consequently there exists  a sequence $\beta$ such that

$$\renewcommand\arraystretch{.4}
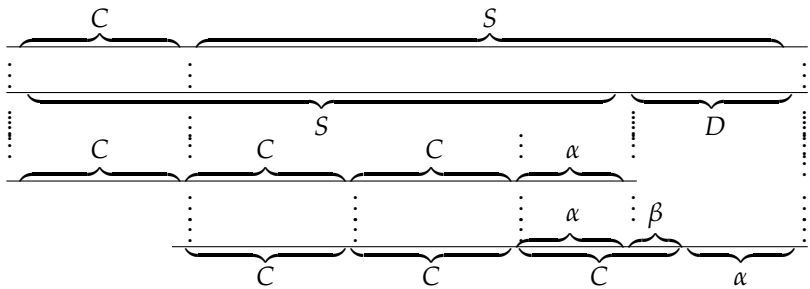
\begin{array}{@{\extracolsep{-8pt}}*{14}{c}}
\jpLetter{3}{C} & \jpLetter{11}{S}\\
\jpOverBrace{3} & \jpOverBrace{11}\\
\hline
\jpLetterRL{3}{} & \jpLetterR{11}{}\\
\hline
\jpUnderBrace{11} & \jpUnderBrace{3}\\[-5pt]
\jpLetterRL{11}{S} & \jpLetterR{3}{D} \\[-15pt]
\jpLetterRL{3}{} & \jpLetterR{8}{} & \jpLetterR{3}{} \\[-10pt]
\jpLetterRL{3}{C} & \jpLetter{3}{C} & \jpLetterR{3}{C} & \jpLetterR{2}{\alpha}& \jpLetterR{3}{}  \\[-10pt]
\jpOverBrace{3} & \jpOverBrace{3} & \jpOverBrace{3} & \jpOverBrace{2}& \jpLetterR{3}{}\\
\cline{1-11}
\jpLetterR{3}{} & \jpLetterR{3}{} & \jpLetterR{3}{}
& \jpLetterR{2}{\alpha}& \jpLetter{1}{\beta}& \jpLetterR{2}{}\\[-10pt]
\jpLetterR{3}{} & \jpLetterR{3}{} & \jpLetterR{3}{}
& \jpOverBrace{2} & \jpOverBrace{1}& \jpLetterR{2}{}\\
\cline{4-14}
\jpLetter{3}{} & \jpUnderBrace{3} & \jpUnderBrace{3}
& \jpUnderBrace{3}& \jpUnderBrace{2}\\[2pt]
\jpLetter{3}{} & \jpLetter{3}{C} & \jpLetter{3}{C}
& \jpLetter{3}{C}& \jpLetter{2}{\alpha}\\
\end{array}$$

\begin{gather*}
C \equiv \alpha \beta \\
\\
D \equiv \beta \alpha \\
\\
S \equiv C^n \alpha \equiv (\alpha \beta)^n \alpha
\equiv \alpha \beta \alpha \beta \alpha \cdots \alpha \beta \alpha
\equiv \alpha (\beta \alpha)^n
\equiv \alpha D^n\\
\\
U \equiv  (\alpha \beta)^{n-1} \alpha \equiv  \alpha (\beta \alpha)^{n-1}
\end{gather*}

\medskip

$C$ is the smallest sequence for which one can find a sequence $D$
such that
$$CS \equiv SD.$$

\medskip

If $\alpha$ contains at least one symbol, then we never have that
$$\alpha \beta \equiv \beta \alpha.$$

Otherwise we would get:
$$\beta S \equiv \beta \alpha (\beta \alpha)^n
\equiv (\beta \alpha)^{n+1}
\equiv (\alpha \beta)^{n+1}
\equiv (\alpha \beta)^{n} \alpha \beta
\equiv S \beta$$
where $\beta$ is composed of fewer symbols than $C$.

\mypage{501}

Each of the sequences
$$\beta \alpha \beta \qquad \hbox{and} \qquad \alpha \beta \alpha$$
where $\alpha$ is composed of at least one symbol, contains only a
single subsequence $\beta \alpha$ and a single subsequence $\alpha \beta$.

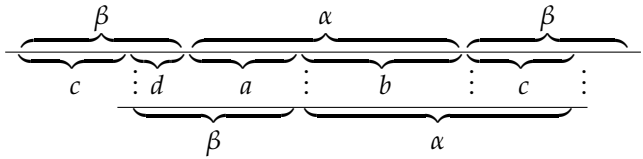
$$\renewcommand\arraystretch{.4}
\begin{array}{@{\extracolsep{-8pt}}*{11}{c}}
\jpLetter{3}{\beta} & \jpLetter{5}{\alpha} & \jpLetter{3}{\beta} \\
\jpOverBrace{3} & \jpOverBrace{5} & \jpOverBrace{3}\\[2pt]
\hline
 \jpUnderBrace{2} &  \jpUnderBrace{1} &  \jpUnderBrace{2} &  \jpUnderBrace{3} &  \jpUnderBrace{2} \\[-5pt]
 \jpLetterR{2}{c} &  \jpLetter{1}{d} &  \jpLetterR{2}{a} &  \jpLetterR{3}{b} &  \jpLetterR{2}{c} \\[3pt]
\cline{3-10}
\jpBlank{2} & \jpUnderBrace{3} & \jpUnderBrace{5}\\[3pt]
\jpBlank{2} & \jpLetter{3}{\beta} & \jpLetter{5}{\alpha}\\
\end{array}$$

If we say that $\beta \alpha \beta$ contains an inner subsequence
$\beta \alpha$, one can write
\begin{gather*}
\alpha \equiv ab \equiv bc\\
\beta \equiv cd \equiv da
\end{gather*}
or
$$\begin{array}{rcl}
S & \equiv &  \alpha \beta \alpha \beta \alpha \cdots \alpha \beta \alpha 
    \equiv (bc)(da)(bc)(da)(bc) \cdots (bc)(da)(bc) \\
  & \equiv & b(cd)(ab)(cd)(ab) \cdots (ab)(cd)(ab)c
    \equiv b \beta[\alpha \beta \alpha \cdots \alpha \beta \alpha c]\\
  & \equiv & [\alpha \beta \alpha \beta \alpha \cdots \alpha c]d\, ab
    \equiv   [\alpha \beta \alpha \beta \alpha \cdots \alpha c]\beta b
\end{array}$$
or 
$$S \equiv b \beta W \equiv W \beta b.$$

However, $b \beta$ here would clearly have to be composed of fewer
symbols than $\alpha \beta$, which is impossible.  In this way it is
also proven that $\beta \alpha \beta$ is not composed of an inner
subsequence $\alpha \beta$.

\medskip

Further, if $\alpha \beta \alpha$ is composed of an inner
subsequence $\beta \alpha$, then we have:

$$\renewcommand\arraystretch{.4}
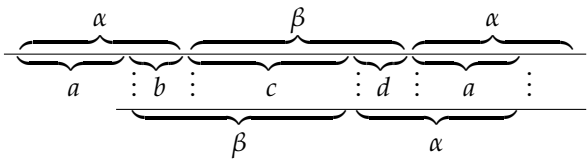
\begin{array}{@{\extracolsep{-8pt}}*{10}{c}}
\jpLetter{3}{\alpha} & \jpLetter{4}{\beta} & \jpLetter{3}{\alpha} \\
\jpOverBrace{3} & \jpOverBrace{4} & \jpOverBrace{3}\\[2pt]
\hline
 \jpUnderBrace{2} &  \jpUnderBrace{1} &  \jpUnderBrace{3} &  \jpUnderBrace{1} &  \jpUnderBrace{2} \\[-5pt]
 \jpLetterR{2}{a} &  \jpLetterR{1}{b} &  \jpLetterR{3}{c} &  
\jpLetterR{1}{d} &  \jpLetterR{2}{a} \\[3pt]
\cline{3-10}
\jpBlank{2} & \jpUnderBrace{4} & \jpUnderBrace{3}\\[3pt]
\jpBlank{2} & \jpLetter{4}{\beta} & \jpLetter{3}{\alpha}\\
\end{array}$$

\begin{gather*}
\alpha \equiv ab \equiv da\\
\beta \equiv bc \equiv cd
\end{gather*}
or
$$\begin{array}{rcl}
S & \equiv & (da)(bc)(da)(bc)(da) \cdots (da)(bc)(da) \\
  & \equiv & d(ab)(cd)(ab)(cd) \cdots (ab)(cd)a\\
  & \equiv & d[\alpha \beta \alpha \beta \alpha \cdots \alpha \beta\, a]\\
  & \equiv & [\alpha \beta \alpha \beta \alpha \cdots \alpha \beta\, a]b\\
\end{array}$$
or 
$$S \equiv d W \equiv W b.$$

That $d$ is composed of fewer symbols than $\alpha \beta$ is however impossible.
In this way it is
also proven that $\alpha \beta \alpha$ is not composed of an inner
subsequence $\beta \alpha$.

\mypage{502}

If one has
$$S \equiv PN \equiv NQ$$
where the number of symbols in $P$ and $Q$ are not less
than\marginnote{So here we're
deliberately selecting some $N$ \emph{shorter} than $U$} the
number in $\alpha \beta$ and $\beta \alpha$, then there is
consequently a whole number $m$ between $0$ and $n$ for which
\marginnote{Thus $P \equiv (\alpha \beta)^{n-m}$ and $Q \equiv
  (\beta \alpha)^{n-m}$} 
$$N \equiv \alpha(\beta \alpha)^m \equiv (\alpha \beta)^m \alpha.$$

In order to find expressions for the sequence $U$ that belongs to the
sequence $S$, we now write:
$$\begin{array}{rcl}
S &\equiv& \alpha_1(\beta_1 \alpha_1)^{n_1} \\
\alpha_1 \beta_1 \alpha_1 &\equiv& \alpha_2(\beta_2 \alpha_2)^{n_2} \\
\hdotsfor{3}\\
\alpha_{p-1} \beta_{p-1} \alpha_{p-1} &\equiv& \alpha_p(\beta_p \alpha_p)^{n_p} \\
\end{array}$$
where $\alpha_{q-1}$ when $2 \leq q \leq p$, contains fewer symbols
than $\beta_q \alpha_q$, while $n_q$ for $1 \leq q \leq p-1$ is
greater than $1$.

Furthermore, let $\alpha_p$ for $n_p > 1$ be completely
missing,\marginnote{When the process terminates, either $n_p=1$ and the
  last factoring is the trivial $\alpha_p \beta_p \alpha_p$, or
  $n_p>1$ and the last factoring is to $\beta_p^{n_p}$.}
while
$\alpha_1(\beta_1 \alpha_1)^{n_1-1}$ is the largest sequence that the
two sequences of $S$, and $\alpha_q(\beta_q \alpha_q)^{n_q-1}$ is the
largest sequence that the two sequences of $\alpha_{q-1} \beta_{q-1}
\alpha_{q-1}$ can have in common. \marginnote{That is, $\alpha_{q-1} \beta_{q-1}
\alpha_{q-1}$ is the `overlap' $U_{q-1}$ on each line}
Depending on whether $n_p$ is greater than or equal to $1$, we can now
treat $\alpha_p$ or $\beta_p$ as $S$ respectively, etc.

\bigskip

If $M$ denotes an arbitrary given sequence, and $N$ denotes the
largest sequence for which one can find sequences $X$ and $Y$ such
that
$$M \equiv XN \equiv NY,$$
then 
$$T \equiv XNY \equiv XM \equiv MY$$
is the shortest sequence for this largest sequence $M$ where
one can find sequences $P$ and $Q$ such that
$$T \equiv PM \equiv MQ.$$
One gets here that
$$P \equiv X, \quad Q \equiv Y.$$

\mypage{503}
\section{V}

Let it be the case that in respect of some null sequence $R$:
\begin{equation}
\left.\begin{array}{rcl}
A_1 & = & B_1\\
A_2 & = & B_2\\
\hdotsfor{3}\\
A_k & = & B_k\\
\end{array}\right\} 
\label{eq:one}
\end{equation}
where two equivalent sequences $A_h$ and $B_h$ for each value of $h$
are composed of the same number of symbols.  One sees immediately that 
 $A_h$ and $B_h$ are composed of equally many\marginnote{Same number
  of $x$s, same number of $y$s etc.  Thus  $A_h$ is just a permutation
  of $B_h$.} of each kind of symbol.

One can write in place of a possible sub-sequence $A_h$ or $B_h$ of
some sequence $S$ the other of these equivalent sequences, so that the
sequence $T$ constructed in this way is equivalent to $S$ in respect
of $R$.  We say that $T$ is constructed from $S$ through a
\emph{homogeneous transformation} according to system \eqref{eq:one}.

Two sequences $S$ and $T$, equivalent in respect of $R$, which are
also equivalent in respect of system \eqref{eq:one} are called \emph{parallel}
sequences in respect of $R$ and \eqref{eq:one}.  We indicate this by writing 
\marginnote{Corollary: Parallel sequences always contain the same number of symbols}
$$S \jpPar T.$$

If two sequences $S$ and $T$ are parallel to one another in respect of
system \eqref{eq:one}, there thus exist such sequences $C_0, C_1, C_2, \cdots,
C_r, C_{r+1}$ where $C_0$ and $C_{r+1}$ denote $S$ and $T$
respectively, so that one can get one of the consecutive sequences $C_m$
and $C_{m+1}$ from the other by exchanging a possible sub-sequence
$A_h$ with the corresponding sequence $B_h$.

When one can not derive any of the equivalences \eqref{eq:one} from the others
through homogeneous transformation we say that the equivalences \eqref{eq:one}
are independent of one another.

\bigskip

Given the sequences
$$zR \hbox{ and } Rz$$
where $R$ denotes the null sequence, if for any symbol $z$ one can
always transform them into one another through homogeneous
transformation by the system \eqref{eq:one}, so that
$$zR \jpPar Rz$$
then we say that \eqref{eq:one} forms a
\emph{complete}\footnote{vollst\"andiges} system of equivalences.

\mypage{504}
Each sub-sequence $R$ of an arbitrary sequence $S$ can thus
through \eqref{eq:one} be moved arbitrarily in the sequence $S$
without changing the order of the remaining symbols of $S$.

In this case we have the following theorem.

\bigskip

If one can get a sequence $\alpha$ from a sequence $A$, and a sequence
$\beta$ from a sequence $B$ by removing a sequence $R$, and meanwhile
one can transform $\alpha$ and $\beta$ into one another by successive
homogeneous transformations according to a complete system of
equivalences, then the sequences $A$ and $B$ have this same property.

We indeed get that e.g.
$$A \jpPar R\alpha \jpPar R\beta \jpPar B.$$

\bigskip

If a system of equivalences derived from a null sequence $R$ has the
property that $A \jpPar B$ whenever $RA \jpPar RB$, then we say that
the system is perfect\footnote{vollkommenes} in respect of $R$.

A complete and perfect system of equivalences in respect of a
null sequence $R$ thus has the property that, in respect of the system
it is always the case that
$$CR \jpPar RC,$$
where $C$ denotes an arbitrary sequence, meanwhile, when
$RA \jpPar RB$ we always have $A \jpPar B$.

\newthought{Theorem.}
If one can get a sequence $\alpha$ from a sequence $A$, and a sequence
$\beta$ from a sequence $B$ by removing a sequence $R$, and meanwhile
in respect of a complete and perfect system of equivalences in respect
of $R$
$$A \jpPar B,$$
then we also have
$$\alpha \jpPar \beta.$$

Then:
$$R\alpha \jpPar A  \jpPar B \jpPar R\beta$$
or
$$\alpha \jpPar \beta.$$

\bigskip

If
$$R \equiv CU \equiv UD,$$
or
$$CR \equiv RD,$$
\mypage{505}
so then in respect of a complete and perfect system of equivalences in respect
of a null sequence  $R$ we always have 
$$C \jpPar D.$$

For
$$RC \jpPar CR \equiv RD.$$

\bigskip

If one has found a complete and perfect system of equivalences in respect
of a null sequence  $R$, then we can immediately see how in this way
our problem \ref{prob:II} is easily solved.

Namely, if $S$ denotes an arbitrary sequence, then one can set up a
series of sequence systems
$$N_0, N_1, N_2, \cdots, N_r$$
that for each value of $p$ all sequences of $N_p$ are parallel, while
$S$ is equal to one of the sequences of $N_0$.  Further the series $N$
can be so chosen that no sequence of $N_r$ contains a sub-sequence $R$,
while it is possible to obtain for any value of $p>0$ a sequence of
$N_{p+1}$ from a sequence of $N_p$ through removal of a sub-sequence
$R$.

Finally, the series $N$ is so chosen that every sequence parallel to a
sequence of the series $N$ is contained in the series $N$.

Having removed then from an arbitrary sequence of a series $N_p$ a
possible sequence $R$, one can obtain in this way for any value of $p
< r$ one of the sequences in the series $N_{p+1}$.

We say now that $N_r$ forms an irreducible sequence system belonging
to $S$.

Our problem \ref{prob:II} is now completed through the remark that similar
sequences, and thus equivalent sequences, must have the same irreducible
sequence system.

For a complete and perfect system of equivalences, a null sequence  $R$
must also be parallel to equivalent sequences with equally many symbols
in respect of the aforementioned system.

We can, however, decide for certain whether or not two sequences are
parallel in a calculable number of steps.

\mypage{506}
\section{VI}

Let there be given the two series of symbol sequences 
\begin{gather*}
A_1, A_2, \cdots, A_k\\
B_1, B_2, \cdots, B_k
\end{gather*}
where $A_p$ and $B_p$ for each value of $p$ are - as before -
corresponding sequences.

Two arbitrary sequences $S$ and $T$ are called equivalent in respect
of the $k$ pairs of corresponding sequences $A_p$ and $B_p$ when there
exist such sequences $C_0,C_1,C_2,\cdots,C_r,C_{r+1}$, where $C_0$ and
$C_{r+1}$ denote $S$ and $T$ respectively, that one can obtain
$C_{q+1}$ from $C_q$ for each value of $q$ through the exchange of a
subsequence $A$ or $B$ for its corresponding sequence.

We represent this, as before, through the equivalence
$$S = T$$
$C_q$ and $C_{q+1}$ are called, as before, equivalent sequences, and
we write
$$C_q \sim C_{q+1}$$.

We have here the equivalences:
\begin{equation}
\left.\begin{array}{c}
A_1 = B_1 \\
A_2 = B_2 \\
\hdotsfor{1}\\
A_k = B_k \\
\end{array}\right\}
\label{eq:two}
\end{equation}

\newthought{Theorem.}
For arbitrary values of $p$ and $q$, let $A_p$ and $B_p$
always start with different symbols on the left, and also for any two
of the sequences in $B$, so we can write:
$$\begin{array}{c}
x_1 P_1 \equiv A_1 = B_1 \equiv y_1Q_1\\
x_2 P_2 \equiv A_2 = B_2 \equiv y_2 Q_2\\
\hdotsfor{1}\\
x_k P_k \equiv A_k = B_k \equiv y_k Q_k,\\
\end{array}$$
where $x$ and $y$ are such symbols that each $y$ is different from
each of the other symbols $y$ and $x$.  If $C$, $M$, $D$ and $N$ are
any such sequences that in respect of system \eqref{eq:two}:
$$CM = DN$$
\mypage{507}
where
$$C = D$$
then it is also the case that
$$M = N.$$

We only need to show here that if
$$zM = zN$$
for any symbol $z$, then $M$ and $N$ must always be equivalent.
For convenience, we will prove the following more comprehensive 
theorem:

Let $zM$ and $zN$, where $z$ denotes a single symbol, be two sequences
that are equivalent in respect of system \eqref{eq:two}, i.e.\ we are given such
sequences $E_1,E_2,\cdots,E_p$ that
\begin{equation}
zM \sim E_1 \sim E_2 \sim \cdots \sim E_p \sim zN,
\label{eq:three}
\end{equation}
then one can find such sequences $F_1,F_2,\cdots,F_q$ that
\begin{equation}
M \sim F_1 \sim F_2 \sim \cdots \sim F_q \sim N, 
\label{eq:four}
\end{equation}
where the number of $F$-sequences $q$ is not greater than the number
of $E$-sequences $p$.

This 
\marginnote{First, three base cases where the derivation is 0, 1 or 2 steps}
theorem is clearly true when
$$zM \equiv zN \quad \hbox{i.e.} \quad M \equiv N.$$

Further, also when
$$zM \sim zN \quad \hbox{i.e.} \quad M \sim N.$$

Finally, 
\marginnote{If this happens we must have applied a rule forwards and then
 backwards, as two different rules must change the first letter.}
the theorem must also be true when 
$$zM \sim E_1 \sim zN,$$
because here $M \equiv N$.

We wish now 
\marginnote{Now three inductive cases...}
to assume that the theorem is true when $1 \leq p < n$.
We will then prove that the theorem is true when $p=n$.
We can then write:
\begin{equation}
zM \sim z_1C_1 \sim z_2C_2 \sim \cdots \sim z_{n-1}C_{n-1} \sim z_nC_n \sim zN,
\label{eq:five}
\end{equation}
where each $z$ denotes a single symbol.
If 
\marginnote{Case 1: If $z$ is
  different from each $x$ and $y$ then the equations
in \eqref{eq:two}  won't allow you to change $z$}
$z$ is different from each $x$ and $y$ one has:
$$z \equiv z_1 \equiv z_2 \equiv \cdots \equiv z_n$$
or
$$M \sim C_1 \sim C_2 \sim \cdots \sim C_{n-1} \sim C_n \sim N.$$

\mypage{508}

If one of the symbols $z_1,z_2,\cdots,z_n$ e.g.\ $z_r$ equals $z$ then
the theorem is also quite clear:
Then we have
\begin{gather*}
zM \sim z_1C_1 \sim z_2C_2 \sim \cdots \sim z_{r-1}C_{r-1} \sim z_rC_r \\
z_rC_r \sim z_{r+1}C_{r+1} \sim z_{r+2}C_{r+2} \sim
\cdots \sim z_nC_n \sim zN, 
\end{gather*}
then 
\marginnote{by the inductive hypothesis}
there exist such sequences $\alpha$ and $\beta$ that
\begin{gather*}
M \sim \alpha_1 \sim \alpha_2 \sim \cdots \sim \alpha_s \sim C_r\\
C_r \sim \beta_1 \sim \beta_2 \sim \cdots \sim \beta_t \sim N,
\end{gather*}
where
$$s+t \leq n-1.$$

We thus need only consider now the case where $z$ denotes and $x$ or a
$y$, while each of the symbols $z_1,z_2,\cdots,z_n$ in \eqref{eq:five} is
different from $z$.

If 
\marginnote{Case 2: start with an $x$}
$z$ were an $x$, e.g.
$$z \equiv x_r$$
then we have in \eqref{eq:five}
$$z_1 \equiv z_2 \equiv \cdots \equiv z_n \equiv y_r$$
$z_1$ in particular being different from $z$, i.e.\
$$z_1 \equiv y_r.$$
If one has further that $z_k \equiv y_r$ while $z_{k+1}$ is different
from $y_r$ then we have
$$z_{k+1} \equiv x_r \equiv z$$
which is clearly impossible.

We thus obtain here
$$x_rM \sim y_rC_1 \sim y_rC_2 \sim \cdots \sim y_rC_n \sim x_rN$$
or
$$x_rP_rM' \sim y_rQ_rM'  \sim \cdots \sim y_rQ_rN' \sim x_rP_rN'$$
where
\begin{gather*}
P_rM' \equiv M\\
P_rN' \equiv N.
\end{gather*}

However, since here 
$$y_rQ_rM'  \sim y_rC_2 \sim \cdots \sim y_rC_{n-1} \sim y_rQ_rN' \sim y_rQ_rN'$$
then one can find such sequences $\gamma$ that
$$M' \sim \gamma_1 \sim \gamma_2 \sim \cdots \sim \gamma_\nu \sim
N',$$
\mypage{509}
where $$\nu \leq n-2,$$
and thus
$$M \equiv P_rM' \sim P_r\gamma_1 \sim  P_r\gamma_2 \sim \cdots \sim  P_r\gamma_\nu \sim
 P_rN' \equiv N.$$

Finally, 
\marginnote{Case 3: start with a $y$; much the same as case 2.}
if  $z$ were equal to a $y$, e.g.\ $y_r$, then we can get from
\eqref{eq:five},
$$y_rM \sim x_rC_1 \sim z_2C_2 \sim \cdots \sim z_{n-1}C_{n-1}
\sim x_rC_n \sim y_rN$$
or
$$y_rQ_rM' \sim x_rP_rM' \sim z_2C_2 \sim \cdots \sim z_{n-1}C_{n-1}
\sim x_rP_rN' \sim y_rQ_rN'$$
where
$$Q_rM' \equiv M, \quad Q_rN' \equiv N.$$

Since
$$x_rP_rM' \sim z_2C_2 \sim \cdots \sim z_{n-1}C_{n-1}
\sim x_rP_rN'$$
then there exists such sequences $\delta$ that
$$M' \sim \delta_1 \sim \delta_2 \sim \cdots \sim \delta_\mu \sim
N',$$
where $$\mu \leq n-2,$$
and thus
$$M \equiv Q_rM' \sim Q_r\delta_1 \sim  Q_r\delta_2 \sim \cdots \sim  Q_r\delta_\mu \sim
 Q_rN' \equiv N.$$

In this way our theorem is proven.

\bigskip

Let $T$ denote an arbitrary sequence such that for each value of a
symbol $z$ it is always the case that
$$zT \equiv Tz.$$

Further, let $T'$ denote an arbitrary sequence equivalent to $T$.
If then 
$$T' \equiv abc \cdots gh,$$
where $a,b,c,\cdots,g,h$ are single symbols, then we would have
$$T' \equiv abc \cdots gh = bc \cdots gha.$$

For
$$a(abc \cdots gh) \equiv aT' = aT = Ta = T'a = (abc \cdots gh)a \equiv a(bc \cdots gha)$$
or
$$abc \cdots gh = bc \cdots gha.$$

Thus if $T$ contains $n$ symbols, then $n$ arbitrary consecutive
symbols of the sequence $TT$ form a sequence equivalent to $T$.

\bigskip

We will now demonstrate some null sequences $R$ for which one can find
a perfect and complete system of equivalences.

\mypage{510}
\section{VII}

\subsection{Example 1.}

Let $R$ be a null sequence defined using the following relations:
\begin{equation}
\left.\begin{array}{r*{2}{@{\;\equiv\;}c}}
R \equiv  X_0 &  (X_1Y_1)^{n_1}X_1 &  X_1(Y_1X_1)^{n_1}\\
   X_1 &  (X_2Y_2)^{n_2}X_2 &  X_2(Y_2X_2)^{n_2}\\
   X_2 &  (X_3Y_3)^{n_3}X_3 & X_3(Y_3X_3)^{n_3}\\
   \hdotsfor{3} \\
   X_{r-1} & (X_rY_r)^{n_r}X_r & X_r(Y_rX_r)^{n_r}\\
\end{array}\right\}
\label{eq:six}
\end{equation}
where $Y_1,Y_2,\cdots,Y_r$ and $X_r$ denote different individual
symbols,\marginnote{These differences are important: 
it means \eqref{eq:seven} will
  then fit the format required for the theorem in \S VI}
while $r$ and each $n$ signify an arbitrary positive whole 
number.

Here we thus have
$$\begin{array}{cc}
R & \equiv (X_1Y_1)^{n_1}X_1 \equiv (X_1Y_1)^{n_1}(X_2Y_2)^{n_2}X_2
\equiv \cdots \\
 & \cdots \equiv (X_1Y_1)^{n_1}(X_2Y_2)^{n_2} \cdots (Y_pX_p)^{n_p}X_p
\equiv \cdots 
\end{array}$$
where
$$1 \leq p \leq r.$$

Further we also have:
$$\begin{array}{cc}
R & \equiv X_1(Y_1X_1)^{n_1} \equiv X_2(Y_2X_2)^{n_2}(Y_1X_1)^{n_1}
\equiv \cdots \\
 & \cdots \equiv X_p(X_pY_p)^{n_p}  \cdots (Y_2X_2)^{n_2}(Y_1X_1)^{n_1}
\end{array}$$
where
$$1 \leq p \leq r.$$

We get now e.g.
$$\begin{array}{c@{\vdots}c@{\vdots}c}
R \equiv X_1Y_1 & (X_1Y_1)^{n_1-1}X_1 & \\
               & (X_1Y_1)^{n_1-1}X_1 & Y_1X_1 \equiv R\\
\end{array}$$
or
$$X_1Y_1 = Y_1X_1.$$

More generally, one obtains for each relevant value of $q > 0$
\begin{gather*}
R \equiv (X_1Y_1)^{n_1} \cdots (X_{q}Y_{q})^{n_{q}} X_{q+1} Y_{q+1}
[ (X_{q+1}Y_{q+1})^{n_{q+1}-1}  X_{q+1} ]\\ ~
[ (X_{q+1}Y_{q+1})^{n_{q+1}-1}  X_{q+1} ]
Y_{q+1} X_{q+1} (Y_{q}X_{q})^{n_{q}}  \cdots  (Y_1X_1)^{n_1} \equiv R
\end{gather*}
or one gets the equivalence:
$$
(X_1Y_1)^{n_1} \cdots (X_{q}Y_{q})^{n_{q}} X_{q+1} Y_{q+1}
=
Y_{q+1} X_{q+1} (Y_{q}X_{q})^{n_{q}}  \cdots  (Y_1X_1)^{n_1}
$$

\mypage{511}
We thus have the following $r$ equivalences in respect of $R$:

\parbox{1.2\textwidth}{\begin{equation}
\left.\begin{array}{c}
X_1Y_1 = Y_1X_1 \\
(X_1Y_1)^{n_1} X_2Y_2 = Y_2X_2 (Y_1X_1)^{n_1}\\
(X_1Y_1)^{n_1} (X_2Y_2)^{n_2} X_3Y_3 = Y_3X_3
(Y_2X_2)^{n_2}(Y_1X_1)^{n_1}\\
\hdotsfor{1} \\
(X_1Y_1)^{n_1} \cdots (X_{r-1}Y_{r-1})^{n_{r-1}} X_rY_r = Y_rX_r
(Y_{r-1}X_{r-1})^{n_{r-1}} \cdots (Y_1X_1)^{n_1}\\
\end{array}\right\}
\label{eq:seven}
\end{equation}
}

We remark here 
\marginnote{and $X_r$ is different from any $Y_i$}
that $X_1$ and $R$ begin on the left with $X_r$.

We add to \eqref{eq:seven} all possible equivalences:
$$R \delta = \delta R$$
where $\delta$ is not equal to any of the symbols $Y_1,Y_2,\cdots,Y_r$
and $X_r$, so the system formed in this way, which we will call $H$,
\marginnote{It's \emph{perfect} because it fits the format required
  for the main theorem of
  \S VI, and we can apply this theorem with $C \equiv D \equiv R$}
is a perfect system.

We will now prove that $H$ is also a complete system in respect of
$R$, or that in respect of $H$ it is always the case that
$$R z \jpPar z R,$$
when $z$ denotes a single arbitrary symbol.

We however need only prove the case where $z$ is equal to one of the
symbols $Y_1,Y_2,\cdots,Y_r$ or $X_r$.

We will however first prove that in respect of $H$ or
\eqref{eq:seven}:

\parbox{1.15\textwidth}{\begin{equation}
(X_1Y_1)^{n_1}\cdots(X_qY_q)^{n_q}[X_{q+1}Y_{q+1}]^m
\jpPar
[Y_{q+1}X_{q+1}]^m(Y_qX_q)^{n_q}\cdots(Y_1X_1)^{n_1}
\label{eq:eight}
\end{equation}}
where $m$ is arbitrary.

\bigskip

The theorem is valid according to \eqref{eq:seven} for $m=0$, $q=1$.
But if the theorem is valid for $q=h$, $m=0$ and for $q=h$, $m=k$, so it is
also valid according to \eqref{eq:seven} for $q=h$, $m=k+1$.

For\marginnote{Some typos in the following:\\
- added superscript $n_h$ in line 2\\
- changed $Y_{h+1}$ to $X_{h+1}$ in line 3
}
$$
\begin{array}{rl}
\multicolumn{2}{l}{(X_1Y_1)^{n_1}\cdots(X_hY_h)^{n_h} [X_{h+1}Y_{h+1}]^{m+1} \equiv} \\
\qquad \equiv & (X_1Y_1)^{n_1}\cdots(X_hY_h)^{\typo{n_h}}[X_{h+1}Y_{h+1}]^{m}X_{h+1}Y_{h+1}
\jpPar \\
\qquad \jpPar &[Y_{h+1}X_{h+1}]^m(Y_hX_h)^{n_h}\cdots(Y_1X_1)^{n_1}\typo{X}_{h+1}Y_{h+1}
\jpPar \\
\qquad \jpPar &[Y_{h+1}X_{h+1}]^m(X_1Y_1)^{n_1}\cdots(X_hY_h)^{n_h}X_{h+1}Y_{h+1}
\jpPar \\
\qquad \jpPar &[Y_{h+1}X_{h+1}]^m Y_{h+1}X_{h+1}(Y_hX_h)^{n_h}\cdots(Y_1X_1)^{n_1}
\equiv \\
& \quad [Y_{h+1}X_{h+1}]^{m+1}(Y_hX_h)^{n_h}\cdots(Y_1X_1)^{n_1}\\
\end{array}$$

Thus \eqref{eq:eight} is also valid for $q=h, m=n$ or for  $q=h+1,
m=0$.

In this way is \eqref{eq:eight} proven.

\mypage{512}
We now get according to \eqref{eq:seven} and \eqref{eq:eight}
\marginnote{Typo: added superscript $n_1$ in the second line}
$$\begin{array}{ccccccc}
Y_q R &\equiv& Y_qX_q(Y_qX_q)^{n_q}\cdots(Y_1X_1)^{n_1} &\jpPar&
(X_1Y_1)^{n_1}\cdots(X_qY_q)^{n_q}X_qY_q &\equiv& RY_q\\
X_r R &\equiv& X_r(X_1Y_1)^{n_1}\cdots(X_rY_r)^{n_r}X_r &\jpPar&
X_r(Y_rX_r)^{n_r}\cdots(Y_1X_1)^{\typo{n_1}}X_r &\equiv& R X_r\\
\end{array}$$

Thus $H$ is also a complete system in respect of $R$.

\bigskip

Our problem \ref{prob:II} is accordingly solved by this means for the given
null sequence $R$.

The above theory keeps its validity if $Y_1,Y_2,\cdots,Y_r$ and
$X_r$ are sequences provided that they cannot overlap with one another.

\subsection{Example 2.}

$$R \equiv ab\, bc\, ab, $$
where $a, b$ and $c$ denote single symbols.

$$\begin{array}{r@{}c@{}l}
R \equiv abbc & [ab] \\
              & [ab] & bcab \equiv R
\end{array}$$
or
$$abbc = bcab$$
or
$$\begin{array}{r@{}c@{}l}
R \equiv abbca & [b] \\
              & [b] & cabab = R
\end{array}$$
or
$$abbca = cabab.$$

In respect of the system\marginnote{As before, this system of
  equivalences fits the format
  required for the main theorem in \S VI, and is thus perfect.}
$$\left.\begin{array}{rcl}
abbc &=& bcab \\
abbca &=& cabab 
\end{array}\right\}$$
we get however
$$\begin{array}{rcl}
aR \equiv & a\,[abbc]\,ab = ab\,[cabab] = ab\,[abbc]\,a = abbcaba & \equiv Ra
\\
& bR \equiv b\,[abbca]\,b = [bcab]\,abb = abbcabb \equiv Rb & \\
& cR \equiv cab\,[bcab] = [cabab]\,bc = abbcabc \equiv Rc.
\end{array}$$

\subsection{Example 3.}

Let
$$R \equiv ABABA,$$
where $A$ and $B$ are such sequences that $ABA$ is the largest
sequence that the two sequences of $R$ have in common.

Further let
$$ABA \equiv UUU\cdots U \equiv U^n$$
where $U$ is not a power sequence, and where $U$ contains more symbols
than $A$.

\mypage{513}

Thus we have here
$$U \equiv AX \equiv YA$$
or
$$B \equiv X U^{n-2} Y.$$

Since $AB = BA$, we get
\begin{gather*}
R = BAABA \equiv XU^{n-2}YAAXU^{n-2}YA \equiv XU^{2n-1} \\
R = ABAAB \equiv AXU^{n-2}YAAXU^{n-2}Y \equiv U^{2n-1}Y
\end{gather*}
or
$$X = Y.$$

If $X$ contains more symbols than $A$, or 
$$U \equiv ACA$$
then we get
$$AC = CA.$$

If $A$ and $C$ here represent single different symbols, or sequences
which cannot have an overlap with each other, then our problem \ref{prob:II} is
solved through these latest equivalences.

\subsection{Example 4.}

$$R \equiv x^nyx^n$$
where $x$ and $y$ are single symbols.

We get
\begin{gather*}
x^ny = yx^n\\
R = x^{2n}y = yx^{2n}\\
\begin{array}{r@{}c@{}l}
R = yx & [x^{2n-1}] & \\
       & [x^{2n-1}] & xy = R
\end{array}
\end{gather*}
or
$$xy = yx$$
which is sufficient.

\subsection{Example 5.}

Let
$$R \equiv x^nyx^nyx^n \ldots x^nyx^n \equiv x^n(y x^n)^p \equiv
(x^ny)^px^n$$
$$n > 1, p >1.$$
Here we have first\marginnote{Thus we can always pull all the $y$'s to
  the left (moving any $x^n$ to the right) or vice versa.}
$$x^ny = yx^n.$$

\mypage{514}
Second one thus gets:
$$\begin{array}{r@{}c@{}l}
R = y^px & [x^{(p+1)n-1}] & \\
       & [x^{(p+1)n-1}] & xy^p = R
\end{array}$$
or
$$y^px = xy^p.$$

\UseKEquationNumbering
We will now show that the equivalences:
\begin{equation}
\left.\begin{array}{c}
x^ny = yx^n\\
y^px = xy^p
\end{array}\right\}
\label{eq:K}
\end{equation}
form a complete and perfect system \ref{eq:K} in respect of $R$.

\bigskip

First 
\marginnote{Proving completeness is the easy part.}
it is certainly\marginnote{Typo: changed $=$ to $\jpPar$ at the
  end of the first equation}
\begin{gather*}
xR \jpPar xx^{(p+1)n}y^p \jpPar x^{(p+1)n}y^px \typo{\jpPar} Rx\\
yR \jpPar yy^px^{(p+1)n} \jpPar y^px^{(p+1)n}y \jpPar Ry.\\
\end{gather*}

\smallskip
Second we will show the following:

\marginnote{This is just the definition of a perfect system being
  spelt out}
If $S$ and $T$ are such sequences that, in respect of the System \ref{eq:K},
$$zS \jpPar zT,$$
so that one can thus find a sequences $E$ where
$$zS \sim E_1 \sim E_2 \sim \cdots \sim E_r \sim zT,$$
where $z$ is an arbitrary symbol denoting $x$ or $y$, then in respect
of  \ref{eq:K} we would also have
$$S \jpPar T.$$

There is then such a sequence $F$ that
$$S \sim F_1 \sim F_2 \sim \cdots \sim F_r \sim T.$$

\marginnote{The proof will be by induction over\\
(a) the length of a derivation and \\
(b) the number of symbols in $S$.}  
Through the figure
$$X \sim Y$$
we will indicate here that one can get $Y$ from $X$ by exchanging a
subsequence $x^ny$ or $yx^n$ or $y^px$ or $xy^p$ for its corresponding
sequence.  

\bigskip

The theorem 
\marginnote{For the base
  cases, note that both the equivalences in  \ref{eq:K} must change the
  leftmost symbol.}
is valid now first when
$zS \sim zT.$
Then clearly
$S \sim T.$

Second the theorem is also valid when
$$zS \sim E \sim zT$$
\mypage{515}
Then clearly
$$S \equiv T.$$

Third, the theorem is valid when both $S$ and $T$ denote just a single symbol.
Then clearly
$$S \equiv T.$$

We assume now \marginnote{Inductive hypothesis}
in advance that the theorem is always true when both
$S$ and $T$ are composed of at most $m$ symbols.
Further we assume that the theorem remains true when both $S$ and $T$
contain $m+1$ symbols, and where the number of $E$-sequences $r$ is
not greater than $n > 1$.

We then need only to prove that the theorem remains true when both $S$ and $T$
are composed of $m+1$ symbols, while in the derivation
$$zS \sim E_1 \sim E_2 \sim \cdots \sim E_r \sim zT$$
the number $r$ of $E$-sequences is equal to $n+1$.

If it is the case that e.g.
$$z \equiv x$$
so we thus have
$$xS \sim z_1C_1 \sim z_2C_2 \sim \cdots \sim z_{n+1}C_{n+1} \sim
xT.$$

If here e.g.
$$z_k \equiv x$$
so we get
$$S \jpPar C_k \jpPar T.$$

In the opposite case\marginnote{i.e. no $z_k$ is $x$} one gets however
$$xS \sim yC_1 \sim yC_2 \sim \cdots \sim yC_{n+1} \sim xT.$$

If here either
$$S \equiv x^{n-1}yS', \qquad T \equiv x^{n-1}yT'$$
or
$$S \equiv y^pS', \qquad T \equiv y^pT'$$
then one gets respectively
\begin{gather*}
xS \equiv x^{n}yS' \sim y(x^nS') \sim \cdots \sim y(x^nT') \sim
x^{n}yT' \equiv xT\\
xS \equiv xy^{p}S' \sim y(y^{p-1}xS') \sim \cdots \sim y(y^{p-1}xT') \sim
xy^{p}T' \equiv xT
\end{gather*}

In both cases we get
$$S' \jpPar T'$$
or
$$S \jpPar T.$$

\mypage{516}

We need then only\marginnote{Since $\sim$ is symmetric this covers the
  other ``two'' cases} to consider the case where e.g.\
$$S \equiv x^{n-1}yS_1, \qquad T \equiv y^pT_1.$$

We get then:
$$xS \equiv x^{n}yS_1 \sim y(x^nS_1) \sim \cdots \sim y(y^{p-1}xT_1) \sim
xy^pT_1 \equiv xT$$
or
$$x^nS_1 \jpPar y^{p-1}xT_1.$$

We get here the alternatives:
\begin{gather*}
y^{p-1}xT_1 \sim \cdots \sim y^{p-1}x(x^{n-1}T_2) \sim \cdots \sim
x^nS_1\\
y^{p-1}xT_1 \sim \cdots \sim y^{p-1}x(y^{p}T_2) \sim y^{2p-1}xT_2 \cdots \sim
x^nS_1.
\end{gather*}

In the first alternative
\begin{gather*}
T_1 \jpPar x^{n-1}T_2\\
x^ny^{p-1}T_2 \jpPar x^nS_1
\end{gather*}
or
$$y^{p-1}T_2 \jpPar S_1$$
or\marginnote{Typo: changed $\equiv$ to $\jpPar$ at the
  end of this equation}
$$S \equiv x^{n-1}yS_1 \jpPar x^{n-1}y^pT_2 \jpPar y^px^{n-1}T_2
\typo{\jpPar} y^pT_1 \equiv T.$$

In the second alternative
$$y^{p-1}xT_1 \sim \cdots \sim y^{qp-1}x(x^{n-1}T_3) \sim  \cdots \sim
x^nS_1$$
or
\begin{gather*}
xT_1 \jpPar y^{p(q-1)}x^nT_3\\
T_1 \jpPar x^{n-1}y^{p(q-1)}T_3 \\
S_1 \jpPar y^{qp-1}T_3
\end{gather*}
or
$$S \equiv x^{n-1}yS_1 \jpPar x^{n-1}y^{qp}T_3 \jpPar y^px^{n-1}y^{(q-1)p}T_3
\jpPar y^pT_1 \equiv T.$$

In this way the theorem is proved.

\section{VIII}
\UseGreekEquationNumbering

Finally we wish to make a few remarks.

If $R$ denotes an arbitrary null sequence, then there exists three
series of symbol sequences
\begin{flalign}
\label{eq:alpha} &P_1, P_2,  \ldots  P_m  \\ 
\label{eq:beta}  &Q_1, Q_2,  \ldots  Q_m  \\ 
\label{eq:gamma} &R_1, R_2,  R_3,  \ldots R_n   
\end{flalign}
with the following properties:
\begin{enumerate}
\item $P_r$ and $Q_r$ are - for each value of $r$ - equivalent to each
  other in respect of $R$, and each of these sequences contains fewer
  symbols than $R$.

\mypage{517}
\item All sequences $R_1, R_2, \ldots R_n$, each of which denote
  a null sequence, are equivalent to one another in respect of the
  equivalences 
\begin{gather}
\left.\begin{array}{c}
P_1 = Q_1\\
P_2 = Q_2\\
\hdotsfor{1}\\
P_m = Q_m
\end{array}\right\}
\label{eq:delta}
\end{gather}

\item For each $r$ the series \eqref{eq:gamma} contains two sequences
  $R_p$ and $R_q$ such that
$$\begin{array}{r@{}c@{}l}
R_p \equiv P_r &U&\\
               &U&Q_r \equiv R_q
\end{array}$$
where $U$ denotes a symbol sequence.

\item If for two arbitrary  sequences
  $R_p$ and $R_q$ of the series \eqref{eq:gamma} 
there exist such symbol sequences $C, D$ and $U$ that
$$\begin{array}{r@{}c@{}l}
R_p \equiv C &U&\\
             &U&D \equiv R_q
\end{array}$$
then the equivalence
$$C = D$$
forms one of the equivalences of \eqref{eq:delta}.
\end{enumerate}

\bigskip

One sees immediately that all the  \eqref{eq:gamma}-sequences contain
equally many symbols, and similarly for $P_r$ and $Q_r$ for each value
of $r$.

\bigskip

We will now show how one can gradually form the sequences in
\eqref{eq:gamma} and the equivalences in \eqref{eq:delta}.

Let 
$$S_1, S_2, \cdots, S_k$$
denote $k$ series of symbol sequences $R$
$$\begin{array}{cccc}
R_1^1, & R_2^1, & \cdots, & R_{n_1}^1\\
R_1^2, & R_2^2, & \cdots, & R_{n_2}^2\\
\hdotsfor{4} \\
R_1^k, & R_2^k, & \cdots, & R_{n_k}^k\\
\end{array}$$
where each $R_x^y$ denotes a single symbol sequence, while
$$R_1^\theta,  R_2^\theta,  \cdots,  R_{n_\theta}^\theta$$
for each $\theta$ is said to denote the series $S_\theta$.

\mypage{518}

Further, we signify by
$$E_1, E_2, \cdots, E_h$$
$h$ systems of equivalences
$$\begin{array}{ccc}
\begin{array}{c}
P^1_1 = Q^1_1\\
P^1_2 = Q^1_2\\
\hdotsfor{1} \\
P^1_{m_1} = Q^1_{m_1}
\end{array}
&
\cdots
&
\begin{array}{c}
P^h_1 = Q^h_1 \\
P^h_2 = Q^h_2 \\
\hdotsfor{1} \\
P^h_{m_h} = Q^h_{m_h} 
\end{array}
\end{array}$$
where each $P$ and each $Q$ denotes a single symbol sequence, and where
$E_\theta$ for each value of $\theta$ is said to represent the system
\begin{gather*}
\begin{array}{c}
P_1^\theta = Q_1^\theta\\
P_2^\theta = Q_2^\theta\\
\hdotsfor{1} \\
P_{m_\theta}^\theta = Q_{m_\theta}^\theta.\\
\end{array}
\end{gather*}

\bigskip

The series $S_1$ only contains the null sequence $R$.

For each value of $\theta$ we form $E_\theta$ from $S_\theta$ and
further $S_{\theta+1}$ from $E_\theta$ as follows:

First, if the system
$$S_1, S_2, \cdots, S_\theta$$
contains two such sequences $R_p$ and $R_q$ that
$$\begin{array}{r@{}c@{}l}
R_p \equiv C &U&\\
             &U&D \equiv R_q
\end{array}$$
where $C, D$ and $U$ denote single symbols or sequences, then
$$C = D$$
is equal to one of the equivalences from $E_\theta$.

For each equivalence
$$P^\theta_r = Q^\theta_r$$
from $E_\theta$ that are opposite for each value of $r$ in the group,
the series $S_1, S_2, \cdots, S_\theta$ contains such sequences $R_p$
and $R_q$ that 
$$\begin{array}{r@{}c@{}l}
R_p \equiv P^\theta_r &U&\\
             &U&Q^\theta_r \equiv R_q
\end{array}$$
\mypage{519}
where  $U$ denotes a single symbol or a sequence.  In this way
$E_\theta$ is completely defined.

Finally, let $S_{\theta+1}$ be formed from all of those unique sequences
$R_{\theta+1}$, that are equivalent to all $R$-sequences in the series
$S_1, S_2, \cdots, S_\theta$ in
respect of the equivalences of the system $E_1, E_2, \cdots,
E_\theta$.

One sees immediately then that $S_\theta$ is contained in
$S_{\theta+1}$ and that $E_\theta$ is contained in $E_{\theta+1}$.

One can however choose $\theta$ so large that
$$S_{\theta+1} \equiv S_\theta$$
and thus also
$$E_{\theta+1} \equiv E_\theta.$$
In this way our claim is proven.

\bigskip

From the system \eqref{eq:delta} we can now choose a system
\eqref{eq:epsilon} of equivalences independent from each other
\begin{gather}
\left.\begin{array}{c}
A_1 = B_1\\
A_2 = B_2\\
\hdotsfor{1}\\
A_k = B_k
\end{array}\right\}
\label{eq:epsilon}
\end{gather}
that one can derive each equivalence in \eqref{eq:delta} from
\eqref{eq:epsilon} while one can thus derive no equivalence in
\eqref{eq:epsilon} from the others.

\eqref{eq:epsilon} can be so chosen that the number $k$ of these
equivalences is minimised.  Further, one can choose \eqref{eq:epsilon}
so that none of these equivalences can be replaced by another with
fewer symbols.

In \eqref{eq:epsilon} it is never the case that
$$A_r \equiv B_r$$
and further we never have simultaneously
\begin{gather*}
A_r \equiv A_s\\
B_r \equiv B_s
\end{gather*}
or
\begin{gather*}
A_r \equiv B_s\\
B_r \equiv A_s.
\end{gather*}

\newthought{Theorem.}
The system \eqref{eq:epsilon} contains no equivalences of
the form
$$TX = TY$$
where e.g.\ $X$ starts the left of one of the sequences $R_x$ of the
\eqref{eq:gamma}-sequences, i.e.
$$XW \equiv R_x.$$
\mypage{520} 

Since the named equivalence must also occur in \eqref{eq:delta}, there
are such sequences $R_y, R_z$ and $R_\mu$ in \eqref{eq:gamma} that:
$$\begin{array}{r@{}c@{}l}
R_y \equiv TX &U&\\
             &U& TY \equiv R_z\\
R_\mu \equiv UT &X&\\
             &X&W \equiv R_x
\end{array}$$
or the equivalence
$$UT = W$$
is contained in \eqref{eq:delta}, or
$$\begin{array}{r@{}c@{}l}
R_x \equiv XW = X &UT&\\
             &UT& Y \equiv R_z.
\end{array}$$

However \eqref{eq:epsilon} then contains the equivalence
$$X = Y,$$
from which one can clearly derive\marginnote{Impossible, since
  by the definition of \eqref{eq:epsilon} you can't derive one of its
  equations   from any of the others.}  
$$TX = TY.$$

\newthought{Theorem.}
The system \eqref{eq:epsilon} contains no equivalence of the
form:
$$SX = SY,$$
where $S$ forms the right end of a  \eqref{eq:gamma}-sequence.

Since the named equivalence must also occur in \eqref{eq:delta}, there
are such sequences $R_x, R_y, R_z$ and $R_\mu$ in \eqref{eq:gamma} that:
$$\begin{array}{r@{}c@{}l}
R_y \equiv SX &U&\\
             &U& SY \equiv R_z\\
R_x \equiv K &S&\\
             &S&XU \equiv R_y
\end{array}$$
or one obtains the equivalence $K = XU$ which is thus contained in
\eqref{eq:delta}.
Finally
$$\begin{array}{r@{}c@{}l}
R_x \equiv KS = X &US& = R_\mu\\
     R_z \equiv \quad &US& Y.
\end{array}$$

However \eqref{eq:delta} then contains the equivalence
$$X = Y,$$
which is impossible.

\mypage{521}
\newthought{Theorem.}
If
$$\begin{array}{p{.4\textwidth}r@{}c@{}lp{.4\textwidth}}
& R_x \equiv P&U\\
and &&&& \\
&             &U&Q \equiv R_y,$$
\end{array}$$
where $R_x$ and $R_y$ denote two sequences from \eqref{eq:gamma}, 
while thus
$$P = Q$$
forms one of the equivalences of  \eqref{eq:delta}, then we have in
respect of \eqref{eq:delta} or in respect of  \eqref{eq:epsilon}
\begin{gather*}
PR = RP\\
QR = RQ\\
UR = RU
\end{gather*}
where $R$ denotes an arbitrary sequence of \eqref{eq:gamma}.

Then
\begin{gather*}
PR = PR_y \equiv PUQ \equiv R_xQ = R_xP = RP\\
UR = UR_x \equiv UPU = UQU \equiv R_yU = RU.
\end{gather*}
Further, one gets in respect of \eqref{eq:epsilon}
\begin{gather*}
PU \equiv R_x = R_y \equiv UQ = UP.
\end{gather*}

\newthought{Theorem.}
If one of the sequences $P$ and $Q$ in \eqref{eq:delta}, which we
shall represent with $C$, has the form
$$C \equiv NM,$$
where $M$ forms the starting left side of a sequence $R_x$ of the
\eqref{eq:gamma}-sequences, then we have for each arbitrary sequence
$R$ of the \eqref{eq:gamma}-sequences in respect of \eqref{eq:delta}:
\begin{gather*}
NR = RN\\
MR = RM.
\end{gather*}

For if $D$ is the sequence corresponding to $C$ in \eqref{eq:delta},
there are clearly such sequences $R_y$ and $R_z$ in \eqref{eq:gamma}
that
$$\begin{array}{r@{}c@{}l}
R_y \equiv NM &U& \\
      &U& D \equiv R_z
\end{array}$$
or\marginnote{Typo: changed equals to equivalence}
$$\begin{array}{r@{}c@{}l}
R_z = UN&M&\\
      &M&W \typo{\equiv} R_x
\end{array}$$
or
$$UN = W$$
or finally
\begin{gather*}
NR = NR_x \equiv NMW = NMUN \equiv R_yN = RN.
\end{gather*}

Further
\begin{gather*}
MR = MR_z \equiv MUNM = MWM \equiv R_xM = RM.
\end{gather*}

\mypage{522}
\newthought{Theorem.}
Let $R_x, R_y, R_z$ and $R_\mu$ denote four arbitrary different or not
different null sequences $R$ of the system \eqref{eq:gamma}.
Further $M$ and $N$ denote two sequences which are obtained through
insertion of $R_z$ and $R_\mu$ in $R_x$ and $R_y$ respectively.  That
is,
\begin{gather*}
M \equiv aR_zb\\
N \equiv c R_\mu d
\end{gather*}
where\marginnote{Typo: changed equals to equivalence}
\begin{gather*}
R_x \equiv ab\\
R_y \;\typo{\equiv}\; cd.
\end{gather*}

If there are then such sequences $C,U$ and $D$ that
$$\begin{array}{r@{}c@{}l}
M \equiv C&U&\\
      &U&D \equiv N
\end{array}$$
then either
$$C = D$$
forms one of the equivalences of \eqref{eq:delta}, or one can obtain
\marginnote{In cases 1-5 below we get $C = D$, while in cases 6-8 $C$ and
  $D$ differ by some $R$.}
sequences from $C$ and $D$ through removal of subsequences $R$ of the
systems \eqref{eq:gamma} that are equivalent in respect of
\eqref{eq:delta}.

Letting the symbol $=$ represent equivalence in respect of
\eqref{eq:delta}, we can distinguish the following
cases:\marginnote{These 8 cases cover all possible configurations
  of the overlap between $aR_zb$ and $c R_\mu d$}

\begin{enumerate}
\item 
$\begin{array}[t]{c}
a \equiv Ce \qquad c \equiv ef\\
b \equiv hi \qquad d \equiv iD\\
\begin{array}{r@{}c@{}l}
R_z \equiv f&g&\\
      &g&h \equiv R_\mu
\end{array}\\
\makebox[\textwidth][l]{or}\\
f = h\\
\begin{array}{r@{}c@{}l}
R_x \equiv Cehi = C&efi&\\
      &efi&D \equiv R_y
\end{array}\\
\makebox[\textwidth][l]{or}\\
C = D.
\end{array}$

\bigskip

\item 
$\begin{array}[t]{c}
a \equiv Ce \qquad c \equiv ef\\
D \equiv hd\\
\begin{array}{r@{}c@{}l}
R_z \equiv f&g&\\
      &g&bh \equiv R_\mu
\end{array}\\
\makebox[\textwidth][l]{or}\\
f = bh\\
\begin{array}{r@{}c@{}l}
R_x \equiv C&eb&\\
      &eb&D \equiv ebhd = efd \equiv R_y
\end{array}\\
\makebox[\textwidth][l]{or}\\
C = D.
\end{array}$

\mypage{523}
\item 
$\begin{array}[t]{c}
a \equiv Cce, \qquad d \equiv gbD\\
\begin{array}{r@{}c@{}l}
R_\mu \equiv e&f&\\
      &f&g \equiv R_z
\end{array}\\
\makebox[\textwidth][l]{or}\\
e = g\\
\begin{array}{r@{}c@{}l}
R_x \equiv Cceb = C&cgb&\\
      &cgb&D \equiv  R_y
\end{array}\\
\makebox[\textwidth][l]{or}\\
C = D.
\end{array}$

\bigskip

\item 
$\begin{array}[t]{c}
C \equiv ae, \qquad D \equiv gd\\
\begin{array}{r@{}c@{}l}
R_z \equiv ec&f&\\
      &f&bg \equiv R_\mu
\end{array}\\
\makebox[\textwidth][l]{or}\\
ec = bg\\
\begin{array}{r@{}c@{}l}
R_x \equiv a&b&\\
      &b&gf = ecf \equiv  R_z
\end{array}\\
\makebox[\textwidth][l]{or}\\
a = gf\\
\begin{array}{r@{}c@{}l}
R_\mu \equiv fbg = fe&c&\\
      &c&d \equiv  R_y
\end{array}\\
\makebox[\textwidth][l]{or}\\
fe = d\\
\makebox[\textwidth][l]{or}\\
C \equiv ae = gfe = gd \equiv D.
\end{array}$

\bigskip

\item 
$\begin{array}[t]{c}
b \equiv gh, \qquad c \equiv fg\\
C \equiv ae, \qquad D \equiv id\\
\begin{array}{r@{}c@{}l}
R_z \equiv e&f&\\
      &f&gd \equiv R_y
\end{array}\\
\makebox[\textwidth][l]{or}\\
e = gd\\
\begin{array}{r@{}c@{}l}
R_x \equiv ag&h&\\
      &h&i \equiv R_\mu
\end{array}\\
\makebox[\textwidth][l]{or}\\
ag = i\\
\makebox[\textwidth][l]{or}\\
C \equiv ae = agd = id \equiv D.
\end{array}$

\bigskip

\item 
$\begin{array}[t]{c}
c \equiv fbg\\
C \equiv ae, \qquad D \equiv g R_\mu d\\
\begin{array}{r@{}c@{}l}
R_z \equiv e&f&\\
      &f&bgd \equiv R_y
\end{array}\\
\makebox[\textwidth][l]{}\\
\end{array}$

\mypage{524}
$\begin{array}[t]{c}
\makebox[\textwidth][l]{or}\\
e = bgd\\
\makebox[\textwidth][l]{or}\\
C \equiv ae = abgd = R_xgd\\
D \equiv gR_\mu d.
\end{array}$

\bigskip

\item 
$\begin{array}[t]{c}
b \equiv ecf\\
C \equiv aR_z e, \qquad D \equiv gd\\
\begin{array}{r@{}c@{}l}
R_z \equiv aec&f&\\
      &f&g \equiv R_\mu
\end{array}\\
\makebox[\textwidth][l]{or}\\
aec = g\\
\makebox[\textwidth][l]{or}\\
D \equiv gd = aecd = ae R_x\\
C \equiv a R_z e.
\end{array}$

\bigskip

\item 
$\begin{array}[t]{c}
b \equiv ef, \qquad c \equiv fg\\
C \equiv aR_x e, \qquad D \equiv g R_\mu d\\
\begin{array}{r@{}c@{}l}
R_x \equiv ae&f&\\
      &f&gd \equiv R_y
\end{array}\\
\makebox[\textwidth][l]{or}\\
ae = gd.
\end{array}$

\end{enumerate}

\bigskip
\noindent
\textsc{i}. May 1914.

\hfill \textsc{\large{\emph{Axel Thue.}}}

\begin{center}
\begin{tabular}{p{.2\textwidth}}
\\ \hline\hline
\end{tabular}
\end{center}


\end{document}